\documentclass[aps,prd,longbibliography,reprint,nofootinbib,amsmath,amssymb,floatfix,superscriptaddress]{revtex4-2}
\usepackage{figures/figures}        
\usepackage{physics}
\usepackage{cleveref}
\usepackage{esint,tensor,isomath}
\usepackage{multirow}
\usepackage{hhline}
\providecommand{\ii}{\text{i}}
\providecommand{\ee}{\text{e}}
\providecommand{\vc}{\vb*}
\providecommand{\uv}{\vu*}
\newcommand{\tens}[1]{\mathsfbfit{{#1}}}
\renewcommand{\Re}{\real}
\renewcommand{\Im}{\imaginary}
\newcommand{\bessel}[1]{\operatorname{J}_{#1}}
\newcommand{\hankel}[2][(1)]{\operatorname{H}^{#1}_{#2}}
\newcommand{\harmol}[1]{\operatorname{Z}_{#1}}

\begin{document}
\title{Optical chiral sorting forces and their manifestation in evanescent waves and nanofibres}
\author{Sebastian \surname{Golat}}                    \email{sebastian.l.golat@kcl.ac.uk}
\affiliation{%
Department of Physics, King's College London, Strand, London WC2R 2LS, United Kingdom}
\author{Jack J. \surname{Kingsley-Smith}}
\affiliation{%
Department of Physics, King's College London, Strand, London WC2R 2LS, United Kingdom}
\affiliation{%
QinetiQ, Cody Technology Park, Farnborough, GU14 0LX, United Kingdom}
\author{Iago \surname{Diez}}
\author{Josep \surname{Martinez-Romeu}}
\author{Alejandro \surname{Martínez}} 
\affiliation{%
Nanophotonics Technology Center, Universitat Politècnica de València, Camino de Vera, s/n Building 8F,
46022, Valencia
Spain}
\author{Francisco J. \surname{Rodr\'iguez-Fortu\~no}} \email{francisco.rodriguez\_fortuno@kcl.ac.uk}
\affiliation{%
Department of Physics, King's College London, Strand, London WC2R 2LS, United Kingdom}

\date{\today}
\begin{abstract}
Optical fields can exert forces of chiral nature on molecules and nanoparticles, which would prove extremely valuable in the separation of enantiomers with pharmaceutical applications, yet it is inherently complex, and the varied frameworks used in the literature further complicate the theoretical understanding. This paper unifies existing approaches used to describe dipolar optical forces and introduces a new symmetry-based `force basis' consisting of twelve vector fields, each weighted by particle-specific coefficients, for a streamlined description of force patterns. The approach is rigorously applied to evanescent waves and dielectric nanofibres, yielding concise analytical expressions for optical forces. Through this, we identify optimal strategies for enantiomer separation, offering invaluable guidance for future experiments.
\end{abstract}

\maketitle%
\section{Introduction}%
The idea that light may exert a force on matter was proposed by Johannes Kepler in the 1600s to explain the observation that a comet’s tail always points away from the sun. Using the phenomenon of optical forces for the levitation and control of small objects in a designed way had to wait many centuries. It was first explored by Arthur Ashkin in the 1970s, who proposed and developed the optical trap \cite{Ashkin1970,Ashkin1971,Ashkin1986} (also called optical tweezers), earning him the 2019 Nobel Prize in Physics. These optical tweezers proved instrumental in the micromanipulation of molecules, cells, viruses, and atoms and have revolutionised research fields with the advent of laser cooling and quantum control of macroscopic objects.

In the optical manipulation of small particles, two main optical forces are considered: the optical gradient, which tends to attract particles towards the maxima in electromagnetic energy, and the optical pressure force, which tends to push particles in the direction of light propagation. Optical tweezers exploit the gradient force, usually stronger for small particles. By tightly focusing a laser beam, a particle is trapped in the high-intensity focus of the beam. The stronger the focusing, the higher the intensity gradient, and the stronger the trapping force---increasing the stiffness of the trap.

Given the success in the manipulation of matter by optical forces, a question has arisen in recent years: can light be used to sort chiral particles and molecules, in other words, to separate two enantiomers (which, apart from being a mirror-reflection of one another are otherwise identical)? This is a challenging problem precisely because two enantiomers are identical in many ways, and because---our body's building blocks being chiral molecules---the opposite enantiomers of a given pharmaceutical molecule may have drastically different effects \cite{Patocka2004,Challener2017,Franks2004}. The answer to the question is positive---if light contains some chirality itself.

The expressions for chiral optical forces have been given in many works \cite{CanaguierDurand2013,Bliokh2014,Yoo2019,Mun2020,Ding2014,Cameron2014,Wang2014,Hayat2015,Chen2016,Zhang2017}, and different proposals for chiral separation of enantiomers using light have been put forth \cite{Ding2014,Cameron2014,Wang2014,Hayat2015,Chen2016,Zhang2017,Tkachenko2014,Cao2018,Zheng2020}, with experimental success found only for chiral microparticles that are large compared to the wavelength \cite{Tkachenko2014}.
A new researcher in this field reading the literature is faced with a difficult task due to the many alternative ways of expressing the forces because of the different notation and symbols, different unit systems, different definitions, and most importantly, different ways of grouping terms that make the expressions look very different from one another. The expressions and theoretical analysis of these chiral optical forces are not as straightforward as the gradient and pressure forces for conventional optical manipulation. There are many terms, and great subtlety is needed. In this work, our aim is threefold: first, to be as clear as possible, going over the main different forms of the expressions used in the literature for chiral optical forces in dipolar particles, pointing out how and why they are equivalent, and making sure they are as general as possible. Second, in the process of clarifying the expressions, we uncover and put forward a novel way of expressing the force, which brings additional clarity, as it highlights the different symmetries involved in the phenomenon. We write the net force as a linear combination of different vector force fields, acting as a basis that describes all possible force patterns, each of them weighted by particle-dependent coefficients. Third, we exemplify how this simplified notation elegantly applies to two simple example cases: an evanescent wave and a cylindrical dielectric nanofibre. In the process, we present generally valid, but surprisingly concise, expressions for diverse electromagnetic quantities in these modes, and based on this formulation, describe the best strategies for enantiomer separation available in these geometries.

This paper is intended to serve as a valuable aid to experimentalists trying to achieve chiral separation of small (dipolar) enantiomers. As such, and contrary to a large fraction of the existing literature, we shall adopt SI units (\emph{Système international d'unités}) together with the convention for writing the speed of light in vacuum as $c_0$, while the phase velocity of electromagnetic radiation in an arbitrary medium will be denoted
$c={\omega}/{k}={c_0}/{n}={1}/{\sqrt{\mu\varepsilon}}$. Throughout the paper, we further assume a dispersion-free, lossless linear background medium and monochromatic fields. We shall denote time-dependent real fields with scripted letters, e.g., $\mathcal{A}(\vb{r}, t)$, while their time-independent phasor representation with regular Latin letters, $A(\vb{r})$, such that $\mathcal{A}(\vb{r}, t) = \Re[A(\vb{r}) \ee^{-\ii \omega t }]$. One of the advantages of this representation is that any time-averaged quantity quadratic in the fields can be obtained by simply
$\expval{\mathcal{A}\mathcal{B}}=\Re[A^*B]$ rather than integrating over time.

We organise this work as follows. In \cref{sec:forces} we review the most general expression for a time-averaged optical force on a small particle. In \cref{sec:dipole} we provide a clear overview of all alternative but equivalent ways of writing this expression for the case of a linear bi-isotropic dipolar particle, and we introduce a new classification of the forces based on the symmetries of the particles and the fields. In the same section, we also introduce the concept of force basis, which can be used to design the fields for separating particles which break certain symmetries and in \cref{sec:evan,sec:fibre} we apply this concept to two simple yet practical examples of evanescent waves and
dielectric nanofibre, respectively.

\section{Optical forces}\label{sec:forces}%
The most infallible tool at our disposal to calculate optical forces is to apply the law of conservation of momentum. Any mechanical momentum that a particle acquires must come at the expense of the reduced electromagnetic momentum in the fields. Therefore, in a steady-state time-harmonic field, the net time-averaged force acting on any piece of matter can be computed as the total electromagnetic momentum entering, per unit of time, any closed volume that includes the object. This is achieved by integrating the momentum flux over a closed surface that surrounds the volume. 
Derived from first principles \cite{Jackson1998,Griffiths2012,Novotny2012}, the time-averaged flux density of electromagnetic momentum is given by the Maxwell stress tensor: 
\begin{equation}\label{eq:mst}
    \!\!\expval{\tens{T}}= \frac{1}{2} \Re \bigg[\varepsilon \vc{E} \otimes \vc{E}^{\ast}\! + \mu \vc{H} \otimes \vc{H}^{\ast}\! - \frac{1}{2}\tens{I} \big(\varepsilon |\vc{E}|^2 + \mu |\vc{H}|^2\big)\bigg],\!\!
\end{equation}
where $\vc{E}$ and $\vc{H}$ are the total electric and magnetic fields, the operator $\otimes$ is the outer (or tensor) product of two vectors, $\tens{I}$ is the $3 \times 3$ identity matrix, an asterisk indicates complex conjugation, and $\varepsilon$ and $\mu$ are the absolute permittivity and permeability of the medium, respectively.

Knowing $\expval{\tens{T}}$, one can compute the net time-averaged force acting on any material body via the flux integral:
\begin{equation}\label{eq:forcefromMST}
    \langle\vc{F}\rangle = \oiint_{\mathcal{S}}^{} \expval{\tens{T}} \dd{\vc{s}},
\end{equation}
where $\dd{\vc{s}}=\uv{n}\dd{s}$ is the surface form, $\uv{n}$ is the outward normal vector of a closed surface $\mathcal{S}$ enclosing the body and $\dd{s}$ is the surface element. We will drop the angular bracket notation for time averages, and hereafter, each observable will be assumed to be time averaged. This general method works on any body, regardless of its size. In this work, we are interested in small particles that undergo Rayleigh scattering and hence behave as dipoles. The criterion for which technique applies on which domain is generally described as follows, where $k = \omega/c = 2\pi/\lambda$ is the medium wavenumber and $a$ is the particle size:
\begin{equation}
	\begin{alignedat}{3}
		ka &\ll{}&& 1 \quad&&{} \text{---}\quad \text{Rayleigh scattering} \\
		ka &\approx{}&& 1 \quad&&{} \text{---}\quad \text{Mie resonant regime} \\
		ka &\gg{}&& 1 \quad&&{} \text{---}\quad \text{geometrical optics} 
	\end{alignedat}
\end{equation}
In the Rayleigh limit, any particle under illumination scatters light in an identical form to an electric $\vc{p}$ and/or a magnetic $\vc{m}$ dipole. Because the analytical expression of the electric and magnetic fields of a general combination of an electric and magnetic dipole is well known, one can substitute such fields into \cref{eq:forcefromMST} and, analytically performing the integration \cite{NietoVesperinas2010, Albaladejo2009, Gao2017}, arrive at the expression for the force acting on a dipole in the Rayleigh limit
\begin{equation}\label{eq:force}
    \!{\vc{F}}=\frac{1}{2}\Re\Big[\!\underbrace{\vphantom{\frac{k^3}{6\pi}}(\grad\!\otimes\!\vc{E})\vc{p}^\ast\!+\!\mu(\grad\!\otimes\!\vc{H})\vc{m}^\ast}_\text{interaction}\underbrace{-\,\frac{k^4\eta}{6\pi}(\vc{p}^\ast\!\!\cp\!\vc{m})}_\text{recoil}\!\Big],\!\!
\end{equation}
where $\vc{p}$ and $\vc{m}$ are electric and magnetic dipole moment vectors, considered as phasors of time-harmonic dipoles, $\vc{E}$ and $\vc{H}$ represent some external applied electric and magnetic fields (excluding those produced by the dipoles) and $\eta=\sqrt{\mu/\varepsilon}$ is the impedance of the surrounding environment.

\Cref{eq:force} constitutes the most general dipole force equation. All the following equations for the force are derived from \cref{eq:force} by assuming the linearity of the particle, such that the induced dipoles are proportional to the applied fields. The linear response of a particle to the applied external fields can be described using complex polarisabilities. These polarisabilities ($\tens\alpha_\text{e},\tens\alpha_\text{m},\tens\alpha_\text{c},\tens\alpha_\text{t}$) are, in general, second-rank tensors that can be represented by $3\times3$ square matrices.
The linear response of dipole moments to external fields can be written as follows:
\begin{equation}
    \label{eq:generaltensorpolarisability} \pmqty{\vc{p}/\sqrt{\varepsilon}\\\sqrt{\mu}\vc{m}}=\pmqty{\tens\alpha_\text{e}&\tens\alpha_\text{t}+\ii\tens\alpha_\text{c}\\\tens\alpha^\intercal_\text{t}-\ii\tens\alpha^\intercal_{\text{c}\vphantom{\text{t}}}&\tens\alpha_\text{m}}\pmqty{\sqrt{\varepsilon}\vc{E}\\\sqrt{\mu}\vc{H}},
\end{equation}
where components of these electric-magnetic vectors are chosen so that they have identical units, e.g., $[\sqrt{\varepsilon}\vc{E}]=[\sqrt{\mu}\vc{H}]=\sqrt{\si{\joule\per\meter\cubed}}$ etc. The literature is not unified on the definitions of these polarisabilities. In particular, different authors choose different units for these polarisabilities by choosing whether they are direct responses to external fields $\vc{E}$ and $\vc{H}$ (used, for example, in \cite{CanaguierDurand2013,Yoo2019,Wang2014,Chen2016,Zhang2017,Hayat2015,Cao2018}) or $\vc{E}$ and $\vc{B}$ (used in \cite{Ding2014,Cameron2014,Chen2016}). 
In this work, following \citet{Mun2020}, we define all polarisabilities so that they all have dimensions of volume, $[\alpha_\text{e}]=[\alpha_\text{m}]=[\alpha_\text{c}]=[\alpha_\text{t}]=\si{\meter\cubed}$, 
making the force expressions more elegant. Polarisabilities are a characteristic of the given particle and typically are proportional to its volume. In this work, we keep the most general form of polarisability, allowing the electric-magnetic coupling terms to have both an antisymmetric part (the chiral polarisability $\alpha_\text{c}$) and a symmetric part (the non-reciprocal polarisability $\alpha_\text{t}$). The latter is usually neglected (\citet{Bliokh2014} includes it but neglects the recoil force) because it only exists for non-reciprocal particles that break time-reversal symmetry. We keep this response not only for the sake of generality, but also because it completes a clear framework relating the symmetries of the particle and the optical fields to the forces that we can expect.

Knowing the incident fields and the polarisability of the particle, the exact calculation of the optical force is straightforward in two steps: (i) calculate the induced dipoles using \cref{eq:generaltensorpolarisability}, and (ii) substitute them into \cref{eq:force}. This substitution may be carried out analytically to arrive at a single-step equation, but the resulting expression is only manageable under certain simplifying assumptions, as discussed in the following section. It is also worth pointing out that if the particle is in an environment that reflects the dipolar fields back into itself, this must be taken into account in both steps. In the first step, the dipole moments must be calculated self-consistently, as the incident fields appearing in \cref{eq:generaltensorpolarisability} have a new term, the self-reflected fields, which depend on the dipole moments themselves. This can also be addressed through the concept of an effective polarisability that depends on the environment \cite{Petrov2015,Sersic2011}. In the second step, the dipolar fields that are reflected back must be included in the fields in the interaction terms of \cref{eq:force}. This complex interaction allows for a plethora of different phenomena to arise, such as repulsion from nearby surfaces \cite{RodriguezFortuno2014,RodriguezFortuno2016,KingsleySmith2020,RodriguezFortuno2018}, pulling forces \cite{Shalin2015,Petrov2015}, self torques \cite{Ginzburg2013}, trapping forces \cite{Ivinskaya2016}, and lateral force recoils due to excitation of nearby guided modes \cite{Wang2014,RodriguezFortuno2015,KingsleySmith2019,Paul2019}. These cases greatly complicate the calculations, but ultimately follow the same physics as included in \cref{eq:generaltensorpolarisability,eq:force}. In the rest of this work, we neglect back-reflection, as if the particle was in a homogeneous background environment.

\section{Force on bi-isotropic dipolar particle}\label{sec:dipole}%
In practice, the polarisability tensors will depend on how the particle is oriented in space relative to the incident fields. However, unless the particles in question are somehow aligned, for instance, by some external static field or by nanofabrication, we can assume that for an average {randomly-oriented} particle within the sample, the effective polarisability will be (bi-)isotropic, and therefore can be described by complex (pseudo)scalars ($\alpha_\text{e},\alpha_\text{m},\alpha_\text{c},\alpha_\text{t}$). In this case \cref{eq:generaltensorpolarisability} becomes
\begin{equation}
\begin{split}
    \vc{p}&=\alpha_\text{e}\varepsilon\vc{E}+(\alpha_\text{t}+{\ii}\alpha_\text{c})\vc{H}/c\,,\\
    \vc{m}&=\alpha_\text{m}\vc{H}+(\alpha_\text{t}-{\ii}\alpha_\text{c})\vc{E}/\eta\,.
\end{split}\label{eq:dipoles}
\end{equation}
Effective polarisabilities for the specific case of a \emph{spherical} chiral bi-isotropic particle can be found in \cref{app:polar}. \Cref{eq:dipoles} can be substituted into \cref{eq:force}, and after separating the real and imaginary parts of polarisabilities and further simplifications, one can write the force in terms of eight local observables that are quadratic in the fields. These quantities are: two real scalars
\begin{equation}\label{eq:scalars}
W_\text{e}=\frac{\varepsilon}{4}\abs{\vc{E}}^2\,,
\quad 
W_\text{m}=\frac{\mu}{4}\abs{\vc{H}}^2\,,
\end{equation}
which represent electric and magnetic energy densities, respectively;
a pair of real pseudovectors
\begin{equation}\label{eq:pseudovecs}
\vc{S}_\text{e}=\frac{1}{4\omega}\Im(\varepsilon{\vc{E}^*\!\cp\vc{E}})\,,
\quad 
\vc{S}_\text{m}=\frac{1}{4\omega}\Im(\mu{\vc{H}^*\!\cp\vc{H}})\,,
\end{equation}
representing electric and magnetic spin angular momentum densities; and a complex pseudoscalar and vector
\begin{equation}\label{eq:mixed}
W_\text{c}=\frac{\ii}{2 c}{\vc{E}^*\!\vdot\vc{H}}\,,\quad 
\vc{\varPi}=\frac{1}{2}{\vc{E}\cp\vc{H}^*}\,,
\end{equation}
where $\Re{W_\text{c}}/\omega=\mathfrak{S}$ is {the} time-averaged helicity density, and $\Re{\vc{\varPi}}$ is the time-averaged Poynting vector representing the flow of active power. Note that quantities $\Im{W_\text{c}}$ and $\Im{\vc{\varPi}}$, while often overlooked, are non-zero in general and also lead to forces felt by particles that break certain symmetries. The vector $\Im{\vc{\varPi}}$ represents the flow of reactive power \cite{Jackson1998}, while the pseudoscalar $\Im{W_\text{c}}$ represents a static ``chirality''\footnote{In the sense of handedness.} due to $\Re\vc{E}\not\perp\Re\vc{H}$ and $\Im\vc{E}\not\perp\Im\vc{H}$, in this sense, 
it can be thought of as a measure of instantaneous \emph{collinearity} of the electric with the magnetic field \cite{Bliokh2014}.

As mentioned before, the resulting force can be written in more than one way. 
One way is to use the complex particle polarisabilities to define the real scalar quantities that are often for their dimensions called cross-sections:%
\begin{equation}\label{eq:crosssection}
\begin{split}
    \sigma_\text{ext}&=\sigma_\text{e}+\sigma_\text{m}=k\Im(\alpha_\text{e}+\alpha_\text{m})\,,\\
    \sigma_\text{rec}&=\frac{k^4}{6 \pi}[\Re(\alpha_{\text{e}}^* \alpha_{\text{m}})+\left|\alpha_{\text{c}}\right|^2-\left|\alpha_\text{t}\right|^2]\,,\\
    \sigma_\text{im}&=\frac{k^4}{6 \pi}[\Im(\alpha_{\text{e}}^* \alpha_{\text{m}})+2 \Re(\alpha_c^* \alpha_\text{t})]\,,\\
    \sigma_\text{re}&=\sigma_\text{ext}-\sigma_\text{rec}\,,
\end{split}
\end{equation}
where $\sigma_\text{ext}$ is indeed an extinction cross-section. For chiral forces, we can also define
pseudoscalar cross-sections:
\begin{equation}\label{eq:pseudosection}
\begin{alignedat}{2}
    \gamma_\text{re}&=k\Im\alpha_\text{c}\,,\quad
    \gamma_\text{im}=k\Im\alpha_\text{t}\\
    \gamma^\text{e}_\text{rec}&=\frac{k^4}{3\pi }[\Re(\alpha_\text{e}^\ast\alpha_\text{c})+\Im(\alpha_\text{e}^\ast\alpha_\text{t})]\\
    \gamma^\text{m}_\text{rec}&=\frac{k^4}{3\pi }[\Re(\alpha_\text{m}^\ast\alpha_\text{c})+\Im(\alpha_\text{m}^\ast\alpha_\text{t})]\\
    \gamma_\text{e}&=2\gamma_\text{re}-\gamma^\text{e}_\text{rec}\,,\quad
    \gamma_\text{m}=2\gamma_\text{re}-\gamma^\text{m}_\text{rec}\,.
\end{alignedat}
\end{equation}%
These can be used to write the force as in \cite{Mun2020,Wang2014,Chen2016,Li2019}
\begin{equation}\label{eq:force1}
    \begin{split}
    \!\!\!\!{\vc{F}}={}&
    \grad{U}+
    \frac{\sigma_\text{re}\Re\vc{\varPi}-\sigma_{\text{im}}\Im\vc\varPi}{c}
    +\omega\qty(\gamma_\text{e} \vc{S}_\text{e}+\gamma_\text{m} \vc{S}_\text{m})\!\!\!\!
    \\&\!\!-\grad\!\cp\!\qty[\frac{\gamma_\text{re}{\Re\vc{\varPi}}\!+\!\gamma_\text{im}{\Im\vc{\varPi}}}{\omega}+c(\sigma_\text{e}\vc{S}_\text{e}\!+\!\sigma_\text{m}\vc{S}_\text{m})
    ]\!\,,\!\!\!\!\!\!\!
\end{split}
\end{equation}
where $U$ is a scalar potential defined as
\begin{equation}
    \!\!\!U=\Re\alpha_\text{e}W_\text{e}+\Re\alpha_\text{m}W_\text{m}+\Re\alpha_\text{c}\Re{W_\text{c}}+\Re\alpha_\text{t}\Im{W_\text{c}}\,.\!\!\!
\end{equation}
The chiral terms of this force are those that are a product of pseudoscalar polarisabilities $\alpha_\text{c}$ and cross sections $\gamma$ with pseudovectors such as $\grad{W_\text{c}}$, $\vc{S}_\text{e}$, $\vc{S}_\text{m}$ and $\curl\vc{\varPi}$. If we reflect either the particle or the fields in a mirror, the pseudoscalars or their corresponding pseudovectors change sign, respectively. The rest of the force is achiral. Some of the literature that uses \cref{eq:force1} gives names to individual terms. The potential $U$ is sometimes called the free energy and $\grad U$ are collectively named the gradient forces. Forces proportional to $\Re\vc{\varPi}$ are referred to as radiation pressure forces\footnote{even though they include the self-recoil of the dipole,} while the term with $\curl\Re\vc{\varPi}$ is known as the vortex force. Spin density forces are the terms directly proportional to $\vc{S}_\text{e}$ and $\vc{S}_\text{m}$, while spin-curl forces are proportional to their curls. The last term to have a name is $\Im\vc{\varPi}$ which is called the flow force because it is related to the alternating flow of ``stored energy'' \cite{Jackson1998}. The term related to the curl of this vector is only present for nonreciprocal particles, it is not included in the literature works and hence does not have a name.

While \cref{eq:force1} is an elegant expression relating kinetic properties of light [\cref{eq:scalars,eq:pseudovecs,eq:mixed}] with observables related to the particle [\cref{eq:crosssection,eq:pseudosection}], it has one disadvantage: the expression no longer allows distinguishing the interaction and recoil parts of the dipole force that appeared in \cref{eq:force}. In order to recover the two terms, we introduce two more 
vectors:
\begin{equation}
\!\!\vc{p}_\text{e}=\frac{\varepsilon}{4\omega}\Im\bqty{(\grad\!\otimes\!\vc{E})\vc{E}^\ast}\,,
\; 
\vc{p}_\text{m}=\frac{\mu}{4\omega}\Im\bqty{(\grad\!\otimes\!\vc{H})\vc{H}^\ast}\,,\!\!
\end{equation}
which are often called canonical (or sometimes orbital) linear momentum densities, on top of which we may define a `complex chiral momentum density' pseudovector:
\begin{equation}
\!\!\vc{p}_\text{c}=\frac{1}{4\omega c}\bqty{(\grad\!\otimes\!\vc{H})\vc{E}^\ast-(\grad\!\otimes\!\vc{E}^\ast)\vc{H}}\,,
\end{equation}
where $\Re\vc{p}_\text{c}$ is the chiral and $\Im\vc{p}_\text{c}$ is the magnetoelectric momentum densities introduced in \citet{Bliokh2014}. These canonical momentum densities are directly related to the previously defined (pseudo)vector observables $\vc{\varPi}, \vc{S}_\text{e}, \vc{S}_\text{m}$ and their curls as follows \cite{Berry2009,Vernon2023}:%
\begin{equation}%
    \begin{split}%
        \vc{p}_\text{e}&=\frac{1}{2c^2}\Re\vc{\varPi}-\frac{1}{2}\curl{\vc{S}_\text{e}}\,,\\
        \vc{p}_\text{m}&=\frac{1}{2c^2}\Re\vc{\varPi}-\frac{1}{2}\curl{\vc{S}_\text{m}}\,,\\
        \Re\vc{p}_\text{c}&=k(\vc{S}_\text{e}+\vc{S}_\text{m})-\frac{1}{2\omega c}\curl{\Re\vc{\varPi}}\,,\\
        \Im\vc{p}_\text{c}&=-\frac{1}{2\omega c}\curl{\Im\vc{\varPi}}\,,
    \end{split}%
\end{equation}%
which, when substituted into \cref{eq:force1}, allows us to rewrite the force in a different way.
Confusion may arise due to the fact that
a fraction of the literature uses \cref{eq:force1} while another uses \cref{eq:force2} below. The two alternative expressions give the same value for the force, but the terms are grouped differently (and some terms in one expression are divided into two terms in the other). While \cref{eq:force1} is simpler in terms of the electromagnetic quantities involved (so we may call it the \emph{EM-centric} formulation), \cref{eq:force2} is simpler in terms of the particle polarisabilities involved (with no cross sections needed, so we may call it the \emph{particle-centric} formulation). This alternative formulation is used by \cite{Yoo2019,Hayat2015,Zhang2017,Cao2018,NietoVesperinas2010} but missing the terms involving $\alpha_\text{t}$. To our knowledge, no previous publication lists both non-reciprocal interaction terms and the recoil force. The full force reads:
\begin{equation}\label{eq:force2}
    \begin{split}
    {\vc{F}}={}&\underbrace{\grad(\Re\alpha_\text{e}W_\text{e}\!+\!\Re\alpha_\text{m}W_\text{m}\!+\!\Re\alpha_\text{c}\Re{W_\text{c}}\!+\!\Re\alpha_\text{t}\Im{W_\text{c}})}_\text{gradient force}
    \\&\underbrace{+2\omega(\Im\alpha_\text{e}\vc{p}_\text{e}\!+\!\Im\alpha_\text{m}\vc{p}_\text{m}\!+\!\Im\alpha_\text{c}\Re\vc{p}_\text{c}\!+\!\Im\alpha_\text{t}\Im\vc{p}_\text{c})}_\text{radiation pressure force}
    \\&\underbrace{-(\sigma_\text{rec}\Re\vc{\varPi}\!+\!\sigma_{\text{im}}\Im\vc\varPi)/c
    -\omega\qty(\gamma^\text{e}_\text{rec} \vc{S}_\text{e}\!+\!\gamma^\text{m}_\text{rec} \vc{S}_\text{m})}_\text{dipole recoil force}.\!\!\!
\end{split}
\end{equation}
where we can see three distinct terms. The first two are generalisations of the conservative gradient and radiation pressure forces, while the third term is purely the recoil force from \cref{eq:force}. The advantage of this form is that it is easier to interpret.  Notice that the gradient forces will always point towards (or away from) the maximal value of \emph{energy densities}, and they depend only on real polarisabilities, which means that a particle will feel it even if there is no absorption of photons, it is a conservative force. The radiation pressure points along the \emph{canonical momenta} of the wave and depends on imaginary polarisabilities, hence extinction cross sections, and it will require photons to be absorbed or scattered. The recoil force points along the flow of \emph{active power} $\Re{\vc\varPi}$, \emph{reactive power} $\Im{\vc\varPi}$ and the electric $\vc{S}_\text{e}$ and magnetic $\vc{S}_\text{m}$ \emph{spin} angular momentum densities and is always proportional to squares of polarisabilities, while the other two terms were linearly proportional to the polarisabilities.

One more formulation is worth mentioning for its simplicity and relationship with symmetries, which was introduced in \citet{Bliokh2014} (although neglecting the recoil force). The key observation is that in free space, Maxwell's equations are symmetric under the
parity inversion (P):
$(x,y,z)^\intercal\mapsto(-x,-y,-z)^\intercal$, time reversal (T):
$t\mapsto-t$,
and the the duality transformation (D):
\begin{equation}
    \qty(\sqrt{\varepsilon}\vc{E},\sqrt{\mu}\vc{H})^\intercal\mapsto\qty(\sqrt{\mu}\vc{H},-\sqrt{\varepsilon}\vc{E})^\intercal,
\end{equation}
which is a symmetry leading to the conservation of optical helicity \cite{Calkin1965}. One can then redefine all the quantities involved in the force such that they are either symmetric or antisymmetric under these transformations. We shall label these quantities with an index $A\in\{0,1,2,3\}$, such that they have the same behaviour under these transformations as Stokes parameters, i.e., $A=0$ will be symmetric under all these, $A=1$ will be antisymmetric under (D), $A=2$ will be antisymmetric under all and $A=3$ will be antisymmetric under parity. The energy densities appearing in the gradient forces can be written as:
\begin{equation}\label{eq:energies}%
    \begin{alignedat}{2}
        W_0&=W_\text{e}+W_\text{m}\,,\quad& W_2&=-\Im{W_\text{c}}\,,\\
        W_1&=W_\text{e}-W_\text{m}\,,\quad& W_3&=\phantom{+}\Re{W_\text{c}}\,,
    \end{alignedat}
\end{equation}%
the canonical momenta featured in the pressure force:
\begin{equation}\label{eq:momenta}%
    \begin{alignedat}{2}
        \vc{p}_0&=\vc{p}_\text{e}+\vc{p}_\text{m}\,,\quad& \vc{p}_2&=-\Im{\vc{p}_\text{c}}\,,\\
        \vc{p}_1&=\vc{p}_\text{e}-\vc{p}_\text{m}\,,\quad& \vc{p}_3&=\phantom{+}\Re{\vc{p}_\text{c}}\,,
    \end{alignedat}
\end{equation}%
and the \emph{spin-like} quantities in the recoil force:
\begin{equation}\label{eq:spins}%
    \begin{alignedat}{2}
        \vc{S}_0&=\vc{S}_\text{e}+\vc{S}_\text{m}\,,\quad& 
        \vc{S}_2&=-\Im{\vc\varPi}/(\omega c)\,,\\
        \vc{S}_1&=\vc{S}_\text{e}-\vc{S}_\text{m}\,,\quad&
        \vc{S}_3&=\phantom{+}\Re{\vc\varPi}/(\omega c)\,.
    \end{alignedat}
\end{equation}%
\begin{table*}[ht!]
    \centering
    \setlength{\tabcolsep}{0pt} 
\renewcommand{\arraystretch}{1.5} 
\begin{tabular}{|cc|cc|cc|c|}
\hline
\multicolumn{2}{|m{2.5cm}|}{\multirow{2}{*}{$\vc{F}=\sum_i \lambda_i \vc{V}\!_i$}\centering} &
  \multicolumn{2}{c|}{\centering Coefficients $\lambda_i$ (dimensions of volume)} &
  \multicolumn{2}{c|}{\centering \hspace{6pt}Basis $\vc{V}\!_i$ (dimensions of force/volume)\hspace{6pt} } &
  \multirow{2}{1.9cm}{\centering symmetries broken by the particle} \\ 
  \hhline{|~~|--|--|~|}
\multicolumn{2}{|c|}{} &
  \multicolumn{1}{c|}{
  \cellcolor{proper}
  \phantom{pseudo}scalar\phantom{pseudo}} &
  \multicolumn{1}{c|}{
  \cellcolor{pseudo}
  pseudoscalar} &
  \multicolumn{1}{c|}{
  \cellcolor{proper}
  \phantom{pseudo}vector\phantom{pseudo}} &
  \multicolumn{1}{c|}{
  \cellcolor{pseudo}
  pseudovector} &
   \\ \hline
\multicolumn{1}{|>{\centering\arraybackslash}p{12pt}|}{\multirow{8}{*}{\rotatebox[origin=r]{90}{Interaction force}}} &
  \multirow{4}{*}{Gradient} &
  \multicolumn{2}{c|}{\cellcolor{proper}
  $\Re(\alpha_{\text{e}}+\alpha_{\text{m}})/2$} &
  \multicolumn{2}{c|}{\cellcolor{proper}$\grad(W_\text{e}+W_\text{m})$} & --
   \\ 
   \hhline{|~~|--|--|-|}
\multicolumn{1}{|c|}{} &
   &
     \multicolumn{2}{c|}{\cellcolor{proper}
     $\Re(\alpha_{\text{e}}-\alpha_{\text{m}})/2$} &
  \multicolumn{2}{c|}{\cellcolor{proper}$\grad(W_\text{e}-W_\text{m})$} & D
   \\ 
   \hhline{|~~|--|--|-|}
\multicolumn{1}{|c|}{} &
   &
   
  \multicolumn{2}{c|}{\cellcolor{pseudo}$\Re\alpha_{\text{c}}$} &
  \multicolumn{2}{c|}{\cellcolor{pseudo}$\grad\Re W_\text{c}$} & P
  
   \\ 
   \hhline{|~~|--|--|-|}
\multicolumn{1}{|c|}{} &
   &
   \multicolumn{2}{c|}{\cellcolor{pseudo}$\Re\alpha_{\text{t}}$} & 
  \multicolumn{2}{c|}{\cellcolor{pseudo}$\grad\Im W_\text{c}$} & T
  \\ 
   \hhline{|~-|--|--|-|} 
\multicolumn{1}{|c|}{} &
  \multirow{4}{*}{Pressure} &
  \multicolumn{2}{c|}{\cellcolor{proper}$\Im(\alpha_{\text{e}}+\alpha_{\text{m}})/2$} &
  \multicolumn{2}{c|}{\cellcolor{proper}$2\omega(\vc{p}_\text{e}+\vc{p}_\text{m})$} & --
   \\ 
   \hhline{|~~|--|--|-|}
\multicolumn{1}{|c|}{} &
   &
   \multicolumn{2}{c|}{\cellcolor{proper}$\Im(\alpha_{\text{e}}-\alpha_{\text{m}})/2$} &
  \multicolumn{2}{c|}{\cellcolor{proper}$2\omega(\vc{p}_\text{e}-\vc{p}_\text{m})$} & D
   \\ 
   \hhline{|~~|--|--|-|}
\multicolumn{1}{|c|}{} &
   &
  \multicolumn{2}{c|}{\cellcolor{pseudo}$\Im\alpha_{\text{c}}$} &
  \multicolumn{2}{c|}{\cellcolor{pseudo}$2\omega\Re\vc{p}_\text{c}$} & P
   \\ 
   \hhline{|~~|--|--|-|}
\multicolumn{1}{|c|}{} &
   &
   \multicolumn{2}{c|}{\cellcolor{pseudo}$\Im\alpha_{\text{t}}$} &
  \multicolumn{2}{c|}{\cellcolor{pseudo}$2\omega\Im\vc{p}_\text{c}$} & T
   \\ \hline
\multicolumn{1}{|c|}{\multirow{4}{*}{\rotatebox[origin=r]{90}{Recoil}}} &
  \multirow{2}{*}{Power} &
  \multicolumn{2}{c|}{\cellcolor{proper}$
    \frac{k^3}{6 \pi}\{\Re[\alpha_{\text{e}}^* \alpha_{\text{m}}]+\abs{\alpha_{\text{c}}}^2-\abs{\alpha_\text{t}}^2\}$} &
  \multicolumn{2}{c|}{\cellcolor{proper}$k\Re{\vc{\varPi}}/c$} & --
   \\  
   \hhline{|~~|--|--|-|} 
\multicolumn{1}{|c|}{} &
   &
     \multicolumn{2}{c|}{\cellcolor{proper}$
    \frac{k^3}{6 \pi}\qty{\Im[\alpha_{\text{e}}^* \alpha_{\text{m}}]+2 \Re[\alpha_c^* \alpha_\text{t}]}$} &
  \multicolumn{2}{c|}{\cellcolor{proper}$k\Im{\vc{\varPi}}/c$} &D
   \\ 
   \hhline{|~-|--|--|-|} 
\multicolumn{1}{|c|}{} &
  \multirow{2}{*}{Spin} &
  \multicolumn{2}{c|}{\cellcolor{pseudo}$\;\;
    \frac{k^3}{6 \pi}\qty{\Re\qty[\alpha_{\text{c}}^*(\alpha_{\text{e}}+\alpha_{\text{m}})]-\Im\qty[\alpha_\text{t}^*(\alpha_{\text{e}}-\alpha_{\text{m}})]}\;\;$} &
  \multicolumn{2}{c|}{\cellcolor{pseudo}$k\omega(\vc{S}_\text{e}+\vc{S}_\text{m})$} & P
   \\ 
   \hhline{|~~|--|--|-|} 
\multicolumn{1}{|c|}{} & 
   &
   \multicolumn{2}{c|}{\cellcolor{pseudo}$\;\;
    \frac{k^3}{6 \pi}\qty{\Re\qty[\alpha_{\text{c}}^*(\alpha_{\text{e}}-\alpha_{\text{m}})]-\Im\qty[\alpha_\text{t}^*(\alpha_{\text{e}}+\alpha_{\text{m}})]}\;\;$} &
  \multicolumn{2}{c|}{\cellcolor{pseudo}$k\omega(\vc{S}_\text{e}-\vc{S}_\text{m})$} & P \& D 
   \\ \hline
\end{tabular}
    \caption{Showing the (pseudo)scalars and (pseudo)vectors involved in the dipolar force and illustrating the idea of a vector basis for the force. The last column lists what symmetry (D for duality, P for parity and T for time reversal) has to be broken by both the particle and the wave for the force component to be non-zero. When the recoil force is negligible, we can say that the force forms an eight-dimensional vector space spanned by basis $\vc{V}\!_i$. Formally, the addition of the recoil terms does not promote the force into a twelve-dimensional vector space because the coefficients of the recoil terms are fully determined by the products of the coefficients of the interaction force.}
    \label{tab:basis}
\end{table*}%
Notably, the quantities in \cref{eq:energies,eq:momenta,eq:spins} are not independent. In particular, the energy densities \cref{eq:energies} are time components of momentum densities \cref{eq:momenta} in the relativistic sense, while the \emph{spin-like} quantities \cref{eq:spins} can be thought of as fluxes of the aforementioned energy densities (for more details, see \cref{app:fluxes}). One can show (see \cref{app:chiraldensities}) that the chiral energy, momentum, and spin densities are simply differences between these quantities carried by the positive/negative helicity components of the electromagnetic field, i.e.,
\begin{equation*}
    W_3=W_+-W_-\,,\quad \vc{p}_3=\vc{p}_+-\vc{p}_-\,,\quad \vc{S}_3=\vc{S}_+-\vc{S}_-\,.
\end{equation*}
This also justifies why they have dimensions of energy, momentum, and spin densities, respectively.

The same procedure  can be followed for the polarisabilities, which leads to the following expressions:
\begin{equation}%
    \begin{alignedat}{2}%
        \alpha_0&=(\alpha_\text{e}+\alpha_\text{m})/2\,,\quad& \alpha_2&=-\alpha_\text{t}\,,\\
        \alpha_1&=(\alpha_\text{e}-\alpha_\text{m})/2\,,\quad& \alpha_3&=\phantom{+}\alpha_\text{c}\,.
    \end{alignedat}%
\end{equation}%
We can also define square polarisabilities $\beta_A$, which are equivalent to the recoil cross sections \cref{eq:crosssection,eq:pseudosection}:
\begin{equation}%
\label{eq:betas}
    \begin{alignedat}{2}
    \beta_0&=
    \frac{k^3}{6 \pi}\qty{\Re\qty[\alpha_{\text{c}}^*(\alpha_{\text{e}}+\alpha_{\text{m}})]-\Im\qty[\alpha_\text{t}^*(\alpha_{\text{e}}-\alpha_{\text{m}})]},
    \\
    \beta_1&=
    \frac{k^3}{6 \pi}\qty{\Re\qty[\alpha_{\text{c}}^*(\alpha_{\text{e}}-\alpha_{\text{m}})]-\Im\qty[\alpha_\text{t}^*(\alpha_{\text{e}}+\alpha_{\text{m}})]},\\
    -\beta_2&=
    \frac{k^3}{6 \pi}\qty{\Im[\alpha_{\text{e}}^* \alpha_{\text{m}}]+2 \Re[\alpha_c^* \alpha_\text{t}]}\\
    \beta_3&=
    \frac{k^3}{6 \pi}\{\Re[\alpha_{\text{e}}^* \alpha_{\text{m}}]+\abs{\alpha_{\text{c}}}^2-\abs{\alpha_\text{t}}^2\}\,
    .
\end{alignedat}
\end{equation}%
The minus signs in front of quantities with the label $A=2$ are chosen to match the signs of Stokes parameters later; however, in \cref{eq:force3}, they will cancel out. Expressed in this \emph{basis}, the optical force takes a surprisingly simple form
\begin{equation}%
    \!\!{\vc{F}}=\sum_{A=0}^3(\underbrace{\Re{\alpha}_A\grad W\!\!_A}_\text{gradient}\underbrace{+2\omega\Im{\alpha}_A\vc{p}\!_A}_\text{pressure}\underbrace{-{\beta}_Ak\omega\vc{S}\!_A}_\text{recoil}),\!\!\label{eq:force3}
\end{equation}%
where {we have included the recoil terms}, which were not previously calculated. While \cref{eq:force1,eq:force2,eq:force3} are only true in the dipole approximation an advantage of 
the symmetry-based approach is that we can always divide the force into components, $\vc{F}=\vc{F}_0+\vc{F}_1+\vc{F}_2+\vc{F}_3$, such that $\vc{F}_0$ is present for every particle and the remaining components $\vc{F}_A$ appear only for particles breaking corresponding symmetries.

We remark that \cref{eq:force1,eq:force2,eq:force3} are all alternative formulations of the \emph{same} force. Ultimately, one may notice that the force on a dipolar particle can be written as a sum of (pseudo)scalar coefficients multiplied by a (pseudo)vector basis. In other words, one can write
\begin{equation}\label{eq:basis_concept}
    \vc{F}=\sum_i \lambda_i \vc{V}\!_i\,,
\end{equation}
where \cref{eq:force1,eq:force2,eq:force3} are simply different choices of the basis vectors and coefficients. For the representation in \cref{eq:force3} we list all the coefficients and the basis in \cref{tab:basis}. Note, however, that this is not formally a vector basis unless we neglect recoil terms, since the coefficients in \cref{eq:betas} are products of the other coefficients. Despite that, the concept illustrated by \cref{eq:basis_concept} is quite profound: it tells us that we have coefficients $\lambda_i$ that carry all the dependence on the particle (and have units of volume), which multiplied by the basis vectors $\vc{V}\!_i$ that depend only on the incident electromagnetic fields (and whose dimensions are force per volume or \emph{force density}) gives us the force that a given particle feels. This can be used in designing optical forces to separate particles with different symmetries. For example, using this idea, one can plot different basis vectors $\vc{V}\!_i(\vc{r})$ for a given electromagnetic mode (as we will do later) to get an idea of the types of forces (the directions and amplitudes) that are possible in a particular waveguide---independently of the particle.

Looking at \cref{tab:basis}, one can see that both the force densities $\vc{V}\!_i$ and associated coefficients $\lambda_i$ have the same behaviour under symmetries of the electromagnetic field, i.e., if one breaks the symmetry the other does as well such that their product (i.e. the force term) is a dual symmetric vector (hence parity odd) which is time reversal even/odd for interaction/recoil force respectively. This gives the formulation in \cref{eq:force3} a very useful interpretation. The terms associated with the total energy $W_0$, the total canonical momentum $\vc{p}_0$, and the active power $\Re{\vc\varPi}$ will be symmetric under D, P, and T, and will be felt by all particles. Other terms will be felt only by particles that break relevant symmetries (see the last column of \cref{tab:basis}). For example, a non-magnetic dielectric particle breaks the D symmetry, so it will also feel D asymmetric forces associated with $W_1=W_\text{e}-W_\text{m}$, $\vc{p}_1=\vc{p}_\text{e}-\vc{p}_\text{m}$ and the reactive power $\Im{\vc\varPi}$. If we had a particle that breaks duality in the opposite sense (a dominant magnetic response), then it would feel these forces in the opposite direction, and one could design an optical field to separate electric and magnetic particles. In our case, we are interested in chiral forces that break the P symmetry, i.e., forces associated with $\Re{W_\text{c}}=W_+-W_-$, $\Re{\vc{p}_\text{c}}=\vc{p}_+-\vc{p}_-$ and spin recoil forces $\vc{S}_\text{e}$ and $\vc{S}_\text{m}$ (note that a dual symmetric chiral particle would only feel $\vc{S}_0=\vc{S}_\text{e}+\vc{S}_\text{m}$). Therefore, these forces are at our disposal if we want to separate enantiomers. As before, each enantiomer will feel a chiral force of the same magnitude but in the opposite direction. As a curiosity, a non-reciprocal particle, i.e., a particle that breaks the T symmetry (as well as D and P), would additionally feel two new force densities $\Im{W_\text{c}}$ and $\Im{\vc{p}_\text{c}}$.

\section{Evanescent wave}\label{sec:evan}%
\begin{figure}[!ht]%
    \centering
    \includegraphics[width=\linewidth]{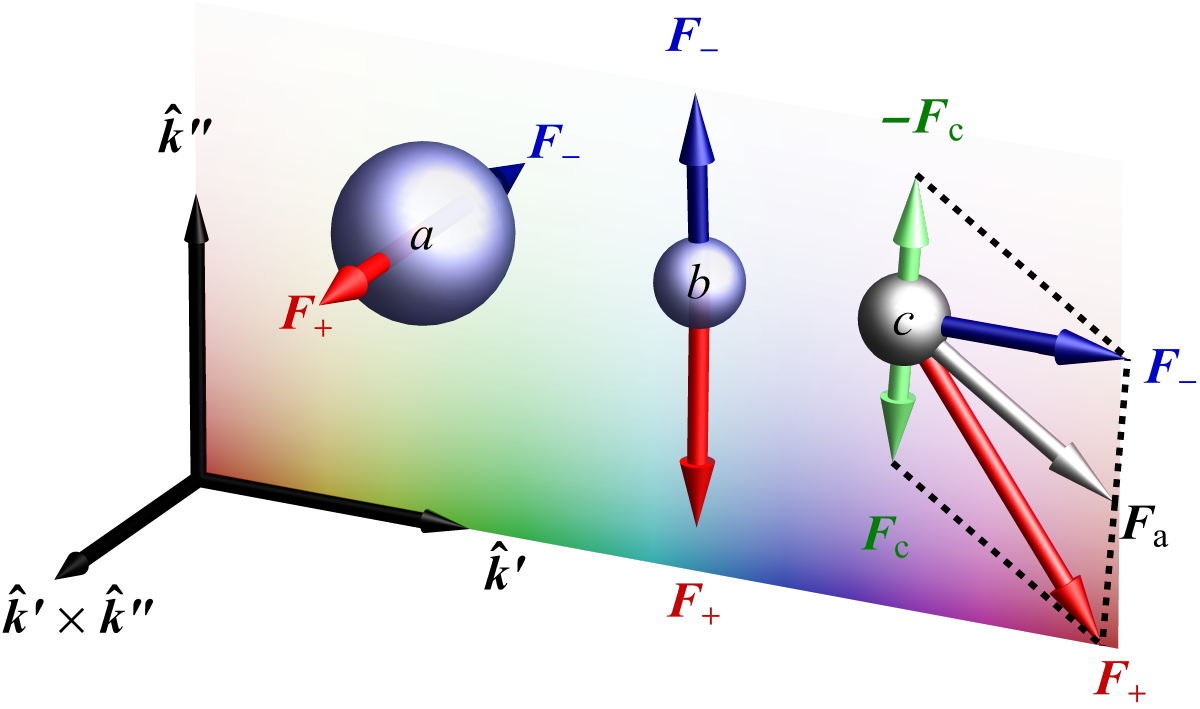}
    \caption{Separation of enantiomers using an evanescent wave. An illustrative figure, forces not to scale. For the larger particle (a) we only show the lateral recoil force, while for the lossless particle (b) and the lossy particle (c) we show the interaction forces.}
    \label{fig:in-plane}
\end{figure}

We may now apply the formalism to simple though practical examples. Consider a particle illuminated by an evanescent wave in a linear lossless medium with a complex wave vector
$\vc{k}=\vc{k}^\prime+\ii\vc{k}^{\prime\prime}=k_p\vu*{k}^\prime+\ii\kappa\vu*{k}^{\prime\prime}$,
where $\kappa>0$ is a real parameter that describes how evanescent the wave is. \Cref{fig:in-plane} illustrates the orthonormal directions of $\vu*{k}^\prime$ (phase advance) and $\vu*{k}^{\prime\prime}$ (decay direction) in an evanescent wave. Because $\vc{k}\cdot\vc{k}=k^2$, we know that $k_p=\sqrt{k^2+\kappa^2}$.
The electric field satisfying the transversality condition $\div\vc{E}=0$ can be written as:
\begin{equation}\label{eq:evan_sol}
    \!\!\sqrt{\varepsilon}\vc{E}(\vc{r})={\ee^{\ii \vc{k}\vdot\vc{r}}}\bqty{A_s\qty(\vu*{k}^{\prime}\!\cp\!\vu*{k}^{\prime\prime})\!+\!A_p\qty(\frac{k_p}{k}\vu*{k}^{\prime\prime}\!\!-\ii \frac{\kappa}{k}\vu*{k}^{\prime})\!},\!\!
\end{equation}
where $A_s$ and $A_p$ are arbitrary complex amplitudes of two allowed polarisations, while the magnetic field can be acquired from the Maxwell–Faraday equation $\curl{\vc{E}}=\ii\omega\mu\vc{H}$. The duality transformation of \cref{eq:evan_sol} can be written as follows:
\begin{equation}\label{eq:duality_evan}
    \sqrt{\varepsilon}\vc{E}\mapsto\sqrt{\mu}\vc{H},\quad(A_s,A_p)\mapsto(A_p,-A_s)\,.
\end{equation}
All observables are quadratic in the fields. Hence, the dependence on polarisation can be written in terms of Stokes parameters, which can be defined in terms of the polarisation amplitudes
\begin{alignat*}{4}
\mathcal{S}_0 &=\abs{A_p}^2+\abs{A_s}^2,
&\quad
\mathcal{S}_2  &=2 \Re(A_p^*A_s)\,, \\
\mathcal{S}_1&=\abs{A_p}^2-\abs{A_s}^2\,,
&\quad
\mathcal{S}_3  &=2 \Im(A_p^*A_s)\,.
\end{alignat*}
Since $A_p$ is a complex scalar and $A_s$ a pseudoscalar, we can say that $\mathcal{S}_0$ and $\mathcal{S}_1$ are scalars, while $\mathcal{S}_2$ and $\mathcal{S}_3$ are pseudoscalars. Additionally, one can confirm using \cref{eq:duality_evan} that $\mathcal{S}_1$ and $\mathcal{S}_2$ are dual asymmetric while the rest are symmetric. 
The energy densities for this field can be calculated using \cref{eq:scalars,eq:energies}, leading to an elegant form, simply proportional to the Stokes parameters: 
\begin{equation}\label{eq:evan_energies}
\begin{alignedat}{4}
    W_0&=\mathcal{S}_0{k_p^2} f(\vc{r})\,,\quad
    &W_2&=\mathcal{S}_2{\kappa^2} f(\vc{r})\,,\\
    W_1&=\mathcal{S}_1{\kappa^2} f(\vc{r})\,,\quad
    &W_3&=\mathcal{S}_3{k_p^2}f(\vc{r})\,,
\end{alignedat}
\end{equation}
where $f(\vc{r})= \exp(-2\vc{k}^{\prime\prime}\!\!\vdot\vc{r})/{2k^2}$. Notice that if a wave is not evanescent $\kappa=0$, then $W_1=W_2=0$, which is directly related to the fact that a plane wave---unlike an evanescent wave---does not break dual symmetry nor time-reversal symmetry. It is also apparent that the parity can be broken even by a circularly polarised plane wave since $W_3$ is proportional to $\mathcal{S}_3$. Knowing the energy densities, remarkably simple equations lead to the energy density gradients $\grad{W}\!_A=-2\vc{k}''{W\!_A}$, and to the canonical momenta $\vc{p}_A=\vc{k}'{W\!_A}/\omega$, as defined in \cref{eq:momenta}. Substituting these into \cref{eq:force3}, we obtain a very simple expression for the interaction force (gradient and pressure terms):
\begin{align}      \label{eq:int_evan} {\vc{F}}_\text{int}&=\sum_{A=0}^32W_A\qty(\Im\alpha_A\vc{k}'-\Re{\alpha}_A\vc{k}'')\,,
\end{align}
which depends only on the energy densities (Stokes's parameters) \cref{eq:evan_energies} and relevant polarisabilities.
The interaction force has no component in the lateral $\vc{k}'\cp\vc{k}''$ direction; forces in this direction are coming purely from the recoil part of the force, which depends on the \emph{spin quantities} \cref{eq:spins}. It is common knowledge that evanescent waves possess transverse spin \cite{Bliokh2014a}, which is independent of helicity ($\mathfrak{S}=W_3/\omega$). The total spin can be calculated from \cref{eq:pseudovecs,eq:spins}, and has two components, longitudinal and transverse:
\begin{equation}
    k\omega\vc{S}_0=\frac{k^2}{k_p^2}W_3\vc{k}'+\underbrace{\frac{k}{k_p^2}W_0\vc{k}'\cp\vc{k}''}_\text{transverse spin}.
\end{equation}
This transverse spin will lead to a chiral lateral recoil force, as used in \citet{Hayat2015}.
It is also known that the real Poynting vector acquires transverse components in the case of a circularly polarised evanescent wave (see, e.g. \citet{Wei2020}) explicitly given by
\begin{equation}
    k\Re(\vc\varPi/c)=k\omega\vc{S}_3=\frac{k^2}{k_p^2}W_0\vc{k}'+\frac{k}{k_p^2}W_3\vc{k}'\cp\vc{k}''.
\end{equation}
Therefore, even an achiral particle will feel an achiral lateral force if irradiated by a circularly polarised evanescent wave, as confirmed experimentally by \citet{Antognozzi2016}.
In addition to transverse spin (contained in $\vc{S}_0$) and transverse Poynting vector (contained in $\vc{S}_3$), the two remaining spin-like quantities from \cref{eq:spins} (which are the reactive power proportinal to $\vc{S}_2$ and the difference in electric and magnetic spins $\vc{S}_1$) will similarly have transverse contributions:
\begin{align}
    k\omega\vc{S}_2&=-\frac{k^2}{\kappa^2}W_1\vc{k}''-\frac{k}{\kappa^2}W_2\vc{k}'\cp\vc{k}''\\
    k\omega\vc{S}_1&=-\frac{k^2}{\kappa^2}W_2\vc{k}''+\frac{k}{\kappa^2}W_1\vc{k}'\cp\vc{k}''
\end{align}
notice that this time, the other contributions point in the direction $\uv{k}''$ rather than $\uv{k}'$. The full recoil force (third term in \cref{eq:force3}) is then
\begin{equation*}
\begin{split}
    \!\!{\vc{F}}_\text{rec}={}&
    k\Big(\frac{{W_2}\beta_2}{\kappa^2}-\frac{{W_0}\beta_0}{k^2_p}+\frac{{W_3}\beta_3}{k^2_p}-\frac{{W_1}\beta_1}{\kappa^2}
    \Big)(\vc{k}'\cp\vc{k}'')
    \\&
    \!\!-\frac{k^2}{k_p^2}\Big(W_3\beta_0
    -W_0\beta_3
    \Big)\vc{k}'
    +\frac{k^2}{\kappa^2}\Big(W_1\beta_2+W_2\beta_1\Big)\vc{k}''. \!\!  
\end{split}
\end{equation*}
The total force $\vc{F} = \vc{F}_\text{int} + \vc{F}_\text{rec}$ exerted by an evanescent wave on a particle can be decomposed into the three directions given by the evanescent wave propagation, decay, and lateral directions. This force depends only on the polarisation of the wave (Stokes' parameters), which controls the energy densities \cref{eq:evan_energies}, and on the evanescence factor $\kappa$, since $k_p=\sqrt{k^2+\kappa^2}$, and is explicitly written as follows:
\begin{equation*}%
    \begin{split}
    \vc{F}\vdot\uv{k}'&=2\sum_{A=0}^3{k_p}W_A\Im\alpha_A+\frac{k^2}{k_p}\Big(W_0\beta_3-W_3\beta_0
    \Big),\\
    \vc{F}\vdot\uv{k}''&=-2\sum_{A=0}^3{\kappa}W_A\Re{\alpha}_A+\frac{k^2}{\kappa}\Big(W_2\beta_1+W_1\beta_2\Big),
\\
    \vc{F}\!\vdot\!\big(\uv{k}'\!\cp\!\uv{k}''\big)\!&=\frac{k{k_p}}{\kappa}\Big({{W_2}\beta_2}\!-\!{{W_1}\beta_1}\Big)\!+\!\frac{k{\kappa}}{k_p}\Big({{W_3}\beta_3}\!-\!{{W_0}\beta_0}\Big).
    \end{split}
\end{equation*}%
The chiral force will only consist of the components that have pseudoscalar coefficients, which (assuming a reciprocal particle, i.e., $\alpha_2=0$) are $\Re\alpha_3 = \Re\alpha_c$, $\Im\alpha_3 = \Im\alpha_c$, $\beta_0={k^3}\Re\left[\alpha_{\text{c}}^*\left(\alpha_{\text{e}}+\alpha_{\text{m}}\right)\right]/({6 \pi})$ and $\beta_1={k^3}\Re\left[\alpha_{\text{c}}^*\left(\alpha_{\text{e}}-\alpha_{\text{m}}\right)\right]/({6 \pi})$. Therefore, the chiral force of a reciprocal particle in an evanescent wave is:
\begin{equation*}%
    \begin{split}
    \vc{F}_\text{\!c}\vdot\uv{k}'&=2{k_p}W_3\Im\alpha_3-\frac{k^2}{k_p}W_3\beta_0
    ,\\
    \vc{F}_\text{\!c}\vdot\uv{k}''&=-2{\kappa}W_3\Re{\alpha}_3+\frac{k^2}{\kappa}W_2\beta_1,
\\
    \vc{F}_\text{\!c}\vdot\big(\uv{k}'\cp\uv{k}''\big)\!&=-\frac{k{k_p}}{\kappa}{{W_1}\beta_1}-\frac{k{\kappa}}{k_p}{{W_0}\beta_0}.
    \end{split}
\end{equation*}%
From this, one can see that in order to maximise the lateral\footnote{Pointing in direction $\uv{k}'\!\cp\!\uv{k}''$.} chiral force, one needs an evanescent wave with a linear polarisation $\mathcal{S}_1=\pm\mathcal{S}_0 $, which also forces non-chiral lateral forces to zero and is precisely
the configuration proposed by \citet{Hayat2015}
and depicted in \cref{fig:in-plane}, particle (a). Note that in that case, the transverse spin-recoil force performs the sorting. Hence, this method will be more successful for larger particles with non-negligible recoil force. 

One cannot perform lateral sorting for particles that are too small to feel recoil forces. If one wants to sort smaller particles that only feel gradient and pressure forces ($\vc{F}_\text{int}$ from \cref{eq:int_evan}), one would have to choose a circular polarisation $\mathcal{S}_3=\sigma\mathcal{S}_0 $ (where $\sigma=\pm1$) and sort the enantiomers in the plane defined by the evanescent wave. This type of sorting of small enantiomers (suitable for chiral molecules) in an evanescent field is a novel proposal, to the best of our knowledge. The circularly polarised evanescent wave has non-zero helicity $W_3\neq0$, but unlike a plane wave, this helicity is not uniform but decays exponentially like the energy. This non-uniformity creates a helicity gradient force that will attract the enantiomer with $\sigma\Re(\alpha_\text{c})>0$ to the region with higher helicity and repel the other. This is shown in \cref{fig:in-plane}, particle (b). The magnitudes of the total interaction force acting on the positive/negative enantiomer (the positive enantiomer is the one with $\Re\alpha_\text{c}>0$) in terms of the chiral polarisability of the positive enantiomer $\alpha_\text{c}$ will be $\abs{\vc{F}_\pm}=2W_0f_\pm$, where $f^2_\pm={k_p^2\Im(\alpha_0\pm\sigma\alpha_\text{c})^2+\kappa^2\Re(\alpha_0\pm\sigma\alpha_\text{c})^2}$.  
For this method, typically the achiral force is stronger than the chiral forces, and so the particles within the evanescent wave will stick to the surface due to gradient forces. But one of the enantiomers feels this force stronger than the other, due to the additive or subtractive chiral force which reinforces or weakens the achiral force. This allows enantiomer separation proposals based on the difference in net gradient forces. If the chiral force is strong enough to overcome the achiral force, then one enantiomer would feel attraction, and the other repulsion, from the surface. This is illustrated in \cref{fig:in-plane}, particle (b).
Particles with non-negligible $\Im\alpha_\text{c}$ will also feel a chiral pressure force in the direction of wave propagation [see \cref{fig:in-plane}, particle (c)], which will introduce an angle $\theta$ between the two forces given by
\begin{equation}
    \cos{\theta}=\sigma\frac{\;k_p^2\;\Im\alpha_\text{c}\Im\alpha_0+\kappa^2\;\Re\alpha_\text{c}\Re\alpha_0\;}{f_+(\kappa,\alpha_0,\alpha_\text{c})f_-(\kappa,\alpha_0,\alpha_\text{c})}.
\end{equation}

\section{Cylindrical nanofibre}\label{sec:fibre}%
\begin{figure}[hb!]
    \centering
    \includegraphics{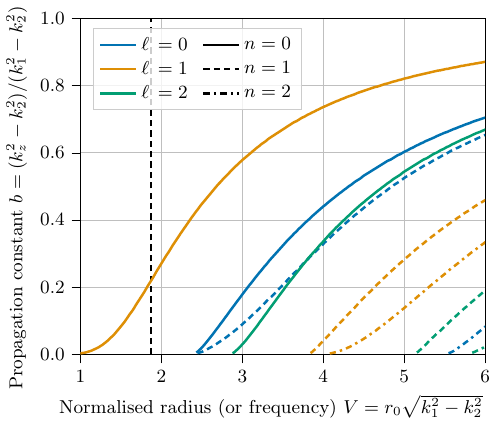}
    \caption{Numerically calculated dispersion relationships for a dielectric fibre made of silicon nitride, $\varepsilon_1=4.3$, immersed in water, $\varepsilon_2=1.7$ (both values are at \SI{20}{\celsius} valid for $\lambda_0$ from $\SIrange{750}{1750}{\nano\meter}$ \cite{rii}). We plot it for the angular momentum eigenmodes $\ell\in\{0,\pm1,\pm2\}$ (in blue, orange, and green). Each $\ell$ has an additional mode number $n$. We plot values $n\in\{0,1,2\}$ (using solid, dashed, and dot-dashed lines). For $\ell\neq0$ the mode number $n$ is related to the spin number of the dominant mode $\sigma=(-1)^n$ and the number of radial nodes in the energy density $n_r=n/2+(\sigma-1)/4$. The orbital angular momentum of the dominant mode is then $\ell-\sigma$. The dashed black line represents the optimal radius (see \cref{eq:opt_rad}), in this case $r_0=\num{0.18}\,\lambda_0$, this value is also used in \cref{fig:circular_forces,fig:linear_forces}.} 
    \label{fig:dispersion}
\end{figure}

\begin{figure*}[!p]
    \centering
    \textbf{Fundamental circular polarisation (CP) mode of the fibre $\ell=1$, $n=0$}
    \par\medskip
    \includegraphics[width=.40\linewidth]{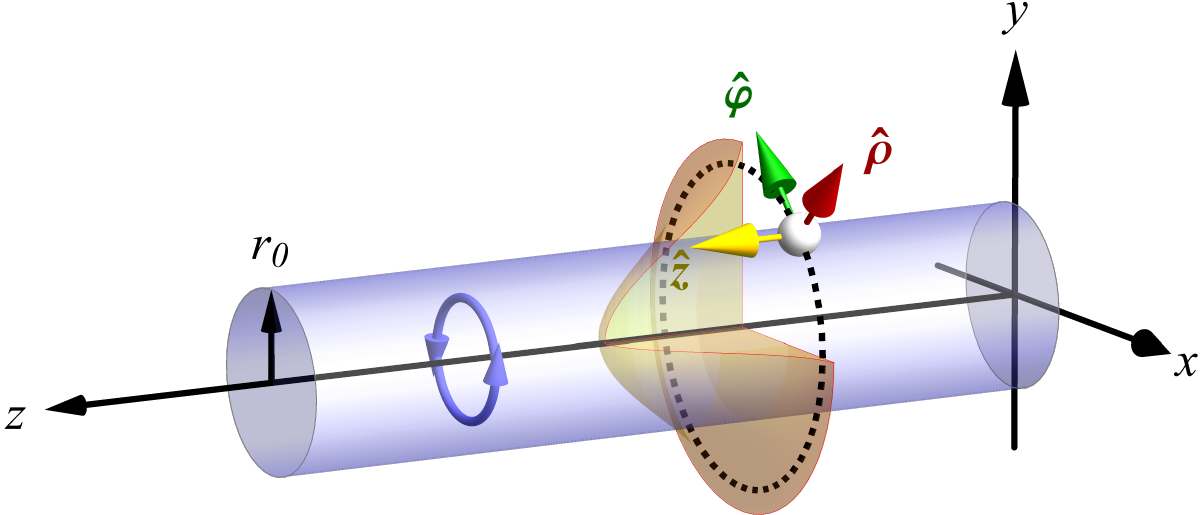}
    \centering
    \includegraphics[width=.96\linewidth]{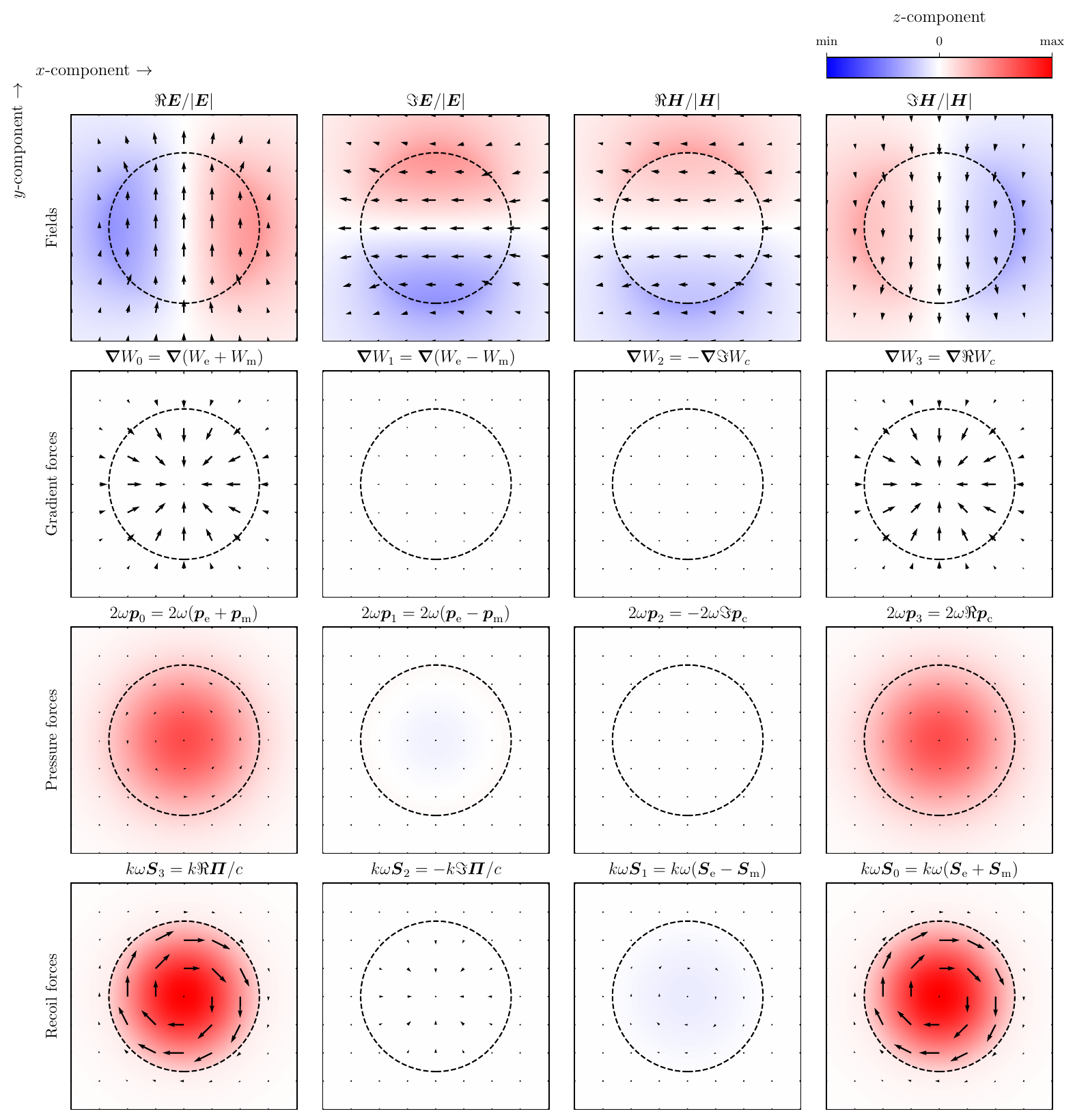}
    \caption{Circular force density basis for optical force around a dielectric fibre guiding total angular momentum eigenmode with $\ell=1$. Arrow maps represent transverse components, while colour maps represent longitudinal components of vector quantities plotted in a cross-sectional view of the fibre. 
    This force basis plot is independent of wavelength. It depends on the material of the fibre SiN ($\varepsilon_1=4.3$), the surrounding medium water ($\varepsilon_2=1.7$) and the radius chosen to be $r_0=\num{0.18}\,\lambda_0$.
    }
    \label{fig:circular_forces}
\end{figure*}

Intuitively, the modal fields outside a nanofibre are equivalent to `radial evanescent waves' and we can expect similar behaviour to the previous section, but with the added richness of having many modes. The eigenmodes of cylindrical waveguides are well studied, since they can be calculated analytically: a particularly concise formulation is provided in \citet{Picardi2018}. A dielectric nanofibre aligned with the $z$-axis can be modelled as a nonmagnetic medium of radius $r_0$, characterized by its permittivity and permeability 
\begin{equation*}
    \varepsilon=\left\{\begin{array}{ll}
\varepsilon_1\varepsilon_0, & \text { for } \rho<r_0 \\
\varepsilon_2\varepsilon_0, & \text { for } \rho>r_0
\end{array} \quad \text { and } \quad \mu=\mu_0\right..
\end{equation*}
The solution to Maxwell’s equations yields electric and magnetic fields that can exist in and around the dielectric fibre. As \citet{Picardi2018} showed, 
it is convenient to choose the spin basis $\uv{e}_{\pm1}=(\uv{x}\pm\ii\uv{y})/\sqrt{2}$ and $\uv{e}_0=\uv{z}$ to represent the solution. On this basis, any vector can be expressed as $\vc{F}=F_{+1}\uv{e}_{+1}+F_0\uv{e}_0+F_{-1}\uv{e}_{-1}$, where the components $F_s$ are \emph{spin-weighted functions}.\footnote{spin-weighted functions have wide use in mathematical physics, particle physics and cosmology. In particular, vector spherical harmonics and Bessel functions can be generalised and simplified by replacing them with their spin-weighted counterparts \cite{TorresdelCastillo2003}.} The advantage of this representation is that rotation around the $z$-axis by an arbitrary angle $\theta$ involves just a phase shift in the spin basis by $s\theta$, $\uv{e}_{s}\mapsto\ee^{\ii s\theta}\uv{e}_{s}$, and a phase shift by $-s\theta$ in the vector components, $F_s\mapsto\ee^{-\ii s\theta}F_s$ \cite{TorresdelCastillo2003}. \citet{Picardi2018} also showed that the total longitudinal angular momentum of the guided modes in the cylindrical fibre is quantised, that is, ${J}_z=\ell W_0/\omega$, where $\ell$ is an integer. The electric field, $\vc{E}$, of a single mode with a definite azimuthal number $\ell$ can be represented by
%
\begin{equation}\label{eq:cyl_sol}
    \begin{split}
    \sqrt{\varepsilon}{E}_{\pm1}(\vc{r})& =\frac{k B\mp\ii k_z A  }{\kappa}\harmol{\ell \mp 1}(\kappa \rho) e^{\ii(\ell\mp1) \varphi+\ii k_z z}, 
    \\
    \sqrt{\varepsilon}{E}_{0}(\vc{r}) & =\sqrt{2} A \harmol{\ell}(\kappa \rho) e^{\ii \ell \varphi+\ii k_z z},
    \end{split}
\end{equation}
where $k$ is the wave number in the medium, $k_z>k_0$ is the mode propagation constant ($k_0$ is the wave number in vacuum), $\kappa=\sqrt{k^2-k_z^2}$ is the radial wave number. Function $\harmol\alpha(x)$ is defined piecewise as the Bessel/Hankel function of the first kind inside/outside of the fibre, respectively. The corresponding magnetic solution, $\vc{H}$, can be obtained via the duality transformation:
\begin{equation}
    \sqrt{\varepsilon}\vc{E}\mapsto\sqrt{\mu}\vc{H},\quad \quad(A, B) \mapsto(B, -A)\,.
\end{equation}
The dispersion relationship for the propagation constant $k_z$ given parameters ($\omega, r_0, \varepsilon_1,\varepsilon_2, \ell$) is found from the transcendental characteristic equation (see \cref{fig:dispersion}), 
whereas the complex constants $A$, $B$ both inside and outside the fibre are determined from the boundary conditions at $\rho=r_0$ (see \cref{app:bondary}). 
Notice that for a mode of a given $\ell$ (denoted by each colour in \cref{fig:dispersion}),
there is more than one dispersion relationship (solid, dashed, and dash-dotted lines). We will label these additional modes with a non-negative integer $n$, which labels the modes from left to right. 
The solution for $\ell=1$, $n=0$ corresponds to a fundamental circularly polarised eigenmode (CP), whose fields can be seen in the first row of \cref{fig:circular_forces}. 
It turns out that the spin-weighted components $F_s$ (either $\sqrt{\varepsilon}E_s$ or $\sqrt{\mu}H_s$ \cref{eq:cyl_sol}), although not being individually physical, are eigenfunctions of definite longitudinal momenta operators with eigenvalues $p_z=\hbar k_z$, $J_z=\hbar\ell$, $S_z=\hbar s$
and $L_z=J_z-S_z=\hbar(\ell-s)$. The momentum operator acts on these eigenfunctions as
\begin{equation}\label{eq:momentum_fibre}
    \uv{p}F_s=-\ii\hbar\grad F_s=\hbar\Bigg({k_z}\uv{z}+\frac{\ell-s}{\rho}\uv{\varphi}-\ii\kappa\frac{\harmol{\ell-s}'}{\harmol{\ell-s}}\uv{\rho}\Bigg)F_s,
\end{equation}
where prime represents the derivative of the function with respect to its argument. Expression in parentheses is a complex wave vector for each component $F_s$
\begin{equation}\label{eq:wavevector}
    \vc{k}_s=\vc{k}_s'+\ii\vc{k}_s''={k_z}\uv{z}+\frac{\ell-s}{\rho}\uv{\varphi}-\ii\kappa\frac{\harmol{\ell-s}'}{\harmol{\ell-s}}\uv{\rho}\,,
\end{equation}
where $\kappa{\harmol{\ell-s}'}/{\harmol{\ell-s}}$ is real both inside and outside the fibre. Notice that the simple longitudinal and azimuthal components of the momentum come from the fact that $F_s$ are eigenfunctions of $p_z$ and $L_z$. In particular, the longitudinal orbital angular momentum operator gives
\begin{equation}
    \hat{L}_zF_s=\uv{z}\vdot(\uv{r}\cp\uv{p})F_s=-\ii\hbar\pdv{\varphi}F_s=\hbar(\ell-s)F_s\,,
\end{equation}
which means that $\uv{z}\vdot(\vc{r}\cp\vc{k}_s)=\ell-s$, which explains the azimuthal component of \cref{eq:wavevector}.\footnote{Recall that in cylindrical coordinates $\vc{r}=z\uv{z}+\rho\uv{\rho}$.}

The orthonormality of the spin basis $\uv{e}_s$ ensures that the scalar product of any two vectors $\vc{F}$ and $\vc{G}$ is
\begin{equation}\label{eq:product}
    \vc{F}^*\!\vdot\vc{G}=F_{+1}^*G^{\vphantom{*}}_{+1}+F_{0}^*G^{\vphantom{*}}_{0}+F_{-1}^*G^{\vphantom{*}}_{-1}\,,
\end{equation}
which means that an eigenmode $\vc{F}$ (either $\sqrt{\varepsilon}\vc{E}$ or $\sqrt{\mu}\vc{H}$) given by $n$ and $\ell$ can be thought of as a linear combination of independent
eigenfunctions $F_s\uv{e}_s$ of definite spin 
\begin{equation}
    \hat{S}_z(F_s\uv{e}_s)=\ii\hbar\uv{z}\cp (F_s\uv{e}_s)=\hbar s (F_s\uv{e}_s)\,.
\end{equation}
\Cref{eq:product} ensures that the energy densities of any eigenmode can be written as a sum
\begin{equation}\label{eq:total_W}
    W\!_A=\sum_{s=-1}^1W\!_A^s=W\!_A^{-1}\!+W\!_A^0+W\!_A^{+1},
\end{equation}
where $W_A^s$ are energy densities calculated for the individual vector components $F_s\uv{e}_s$ if all other components were zero.
To calculate these, one can define two real and one complex coefficient quadratic in scalar $A$ and pseudoscalar $B$ such that they have the same symmetries as $W_0$, $W_3$ and $W_\text{c}$:
\begin{alignat*}{2}
    w_\pm&=
    \frac{1}{4}\pqty{{\abs{A}^2\pm\abs{B}^2}}\,,\quad&
    w_\text{c}&=
    \frac{\ii}{2} A^* B.
\end{alignat*}
One can then write the energy densities [\cref{eq:energies}] associated with each spin-weighted component. The transverse spin ($s=0$) component has
\begin{equation}\label{eq:long_energies}
\begin{alignedat}{4}
    W^0_0&=2w_+\abs{\harmol{\ell}}^2\,,\quad
    -&W^0_2&=2\Im(w_\text{c})\abs{\harmol{\ell}}^2\,,\\
    W^0_1&=2w_-\abs{\harmol{\ell}}^2\,,\quad
    &W^0_3&=2\Re(w_\text{c})\abs{\harmol{\ell}}^2\,,
\end{alignedat}
\end{equation}
while the remaining components will be mixed
\begin{equation}\label{eq:spin_energies}
\begin{alignedat}{4}
    \!\!W^{\pm1}_0&=\bqty{\frac{k_z^2+k^2}{\abs{\kappa}^2}W_0^0\pm\frac{2kk_z}{\abs{\kappa}^2}W_3^0}\frac{\abs{\harmol{\ell\mp1}}^2}{2\abs{\harmol{\ell}}^2},\!\!\\
    W^{\pm1}_1&=\bqty{{\frac{k_z^2-k^2}{\abs{\kappa}^2}}W^0_1}\frac{\abs{\harmol{\ell\mp1}}^2}{2\abs{\harmol{\ell}}^2},\!\!\\
    W^{\pm1}_2&=\bqty{\frac{k_z^2-k^2}{\abs{\kappa}^2}W^0_2}\frac{\abs{\harmol{\ell\mp1}}^2}{2\abs{\harmol{\ell}}^2},\\
    \!\!W^{\pm1}_3&=\bqty{\frac{k_z^2+k^2}{\abs{\kappa}^2}W_3^0\mp \frac{2kk_z}{\abs{\kappa}^2}W_0^0}\frac{\abs{\harmol{\ell\mp1}}^2}{2\abs{\harmol{\ell}}^2},
\end{alignedat}
\end{equation}
where $k_z$ changes sign under a parity transformation.
For a circular polarisation, from the symmetries, one would expect that $w_-$ and $\Im w_\text{c}$ will vanish, but that is not the case for a dielectric fibre due to the fact that there is electromagnetic asymmetry $A\neq\pm\ii B$. This asymmetry comes from the refractive index being due only to the permittivity $\varepsilon_1$. If instead we have a fibre made of material that has $\varepsilon_1=\mu_1=n^2$, then $w_-$ and $\Im w_\text{c}$ will vanish.

By decomposing the mode into spin-weighted functions, we may write the energy density gradients identically to a sum of three evanescent waves with $\grad{W^s_A}=-2\vc{k}_s''W^s_A$, each having wavevector given by \cref{eq:wavevector} and energy densities given by \cref{eq:long_energies,eq:spin_energies}:
\begin{align}\label{eq:fibre_grad}
    \!\!\grad W_A&=\sum_{s=-1}^{1}\!\!2{W^{s}_A}\kappa\frac{\harmol{\ell-s}'}{\harmol{\ell-s}}\uv{\rho}.
\end{align}
This is valid for all energy density quantities $A=0,1,2,3$ defined in \cref{eq:energies}. Remarkably, the same holds for canonical momenta, $\vc{p}^s_A=\vc{k}_s'{W\!\!_A}/\omega$, leading to
\begin{align}\label{eq:fibre_momenta}
    \vc{p}_A=\sum_{s=-1}^{1}\!\!\frac{W^{s}_A}{\omega}\bqty{{k_z}\uv{z}+\frac{\ell-s}{\rho}\uv{\varphi}}.
\end{align} 

Notice that if we introduced factors of $\hbar$ in both the numerator and denominator in the above sums, then the canonical momenta (gradients) would look like a weighted average of real (imaginary) momentum carried by each spin-weighted function
\begin{equation*}
    \vc{p}_A=\sum_{s=-1}^{1}\!\!n^s_A\,\hbar\vc{k}'_s\,,\quad \grad{W\!\!_A}/(2\omega)=-\sum_{s=-1}^{1}\!\!n^s_A\,\hbar\vc{k}''_s\,,
\end{equation*}
where weights $n^s_A={W^{s}_A}/{(\hbar\omega)}$ for $A=0$ can be intuitively thought of as the number density of photons with the momentum given by \cref{eq:momentum_fibre}. For $A=3$, it would be the number density of these photons that break parity, duality for $A=1$, or both and time reversal for $A=2$. Notice that for $A\neq0$, these number densities can be positive or negative, representing whether the photons are more right/left-handed, electric/magnetic, etc. Using the same logic, \cref{eq:total_W} is a weighted average of the energies $\hbar\omega$ carried by each spin component.

The interaction force for a pure fibre eigenmode (later we study mode combinations) will then be a sum of forces identical to three evanescent waves \cref{eq:int_evan} with wavevector given by \cref{eq:wavevector} for each of the spin components
\begin{equation}      \label{eq:int_fibre} {\vc{F}}_\text{int}=\sum_{A=0}^3\sum_{s=-1}^{1}2W^s_A\qty(\Im\alpha_A\vc{k}_s'-\Re{\alpha}_A\vc{k}_s'')\,.
\end{equation}
Unfortunately, this is where the similarities with the evanescent waves end. If a particle is big enough (with respect to the wavelength), it will also feel the recoil force, which depends on the fluxes of power and helicity. Those are related to the spin operator, but we can only repeat the trick we used in \cref{eq:momentum_fibre} for the $z$ component. 
Therefore, the spin densities have to be obtained from the definition \cref{eq:spins}. The total spin density will be
\begin{equation}
    \!\!\vc{S}_0=\sum_{s=-1}^{1}\frac{W_0^{s}}{\omega}s\uv{z}-2\qty(\frac{W_0^{0}}{\omega} \frac{k_z}{\kappa} \frac{Z_\ell^{\prime}}{Z_\ell}- \frac{W_3^{0}}{\omega}\frac{k\ell}{\kappa^2 \rho})\uv{\varphi},\!\!
\end{equation}
where we can see that the longitudinal component is again just the weighted average of the eigenvalues of each spin component. Notice that there is no longitudinal contribution from the eigenfunction with $s=0$. Instead, this function will lead to a transverse spin, which depends only on $W_A^0$ and $\vc{k}_0$. Notice that when we talk about longitudinal and transverse spin in this scenario, we mean with respect to the $z$-direction along the fibre. 
The spin quantity related to the real Poynting vector $\omega\vc{S}_3=\Re\vc{\varPi}/c$ will be very similar:
\begin{equation}
    \!\!\vc{S}_3=\sum_{s=-1}^{1}\frac{W_3^{s}}{\omega}s\uv{z}-2\qty(\frac{W_3^{0}}{\omega} \frac{k_z}{\kappa} \frac{Z_\ell^{\prime}}{Z_\ell}+ \frac{W_0^{0}}{\omega}\frac{k\ell}{\kappa^2 \rho})\uv{\varphi},\!\!
\end{equation}
and so will be the spin asymmetry vector
\begin{equation}
    \!\!\vc{S}_1=\sum_{s=-1}^{1}\frac{W_1^{s}}{\omega}s\uv{z}-2\qty(\frac{W_1^{0}}{\omega} \frac{k_z}{\kappa} \frac{Z_\ell^{\prime}}{Z_\ell}- \frac{W_2^{0}}{\omega}\frac{k\ell}{\kappa^2 \rho})\uv{\varphi}.\!\!
\end{equation}
One can see that all of these quantities have only longitudinal and azimuthal components; the only exception to this will be the spin quantity related to the imaginary Poynting vector $\omega\vc{S}_2=-\Im\vc{\varPi}/c$
\begin{equation}
    \!\!\vc{S}_2=\sum_{s=-1}^{1}\frac{W_2^{s}}{\omega}s\uv{z}+\frac{2}{\kappa} \frac{Z_\ell^{\prime}}{Z_\ell}\qty(k_z\frac{W_2^{0}}{\omega}\uv{\varphi}-k\frac{W_1^{0}}{\omega}\uv{\rho}),\!\!
\end{equation}
which also has a radial contribution.
The analytical form of the total force $\vc{F} = \vc{F}_\text{int} + \vc{F}_\text{rec}$ on a particle near the fibre is then, in cylindrical basis:
\begin{equation*}%
    \begin{split}
    \!\!F_{\rho}=&\sum_{A,s}2\kappa\frac{\harmol{\ell-s}'}{\harmol{\ell-s}}\Re\alpha_A{W^{s}_A}-\frac{2k^2}{\kappa}\frac{Z_\ell^{\prime}}{Z_\ell}\beta_2W_1^{0}
    ,\!\!\\
    \!\!F_{\varphi}=&\sum_{A,s}2\frac{\ell-s}{\rho}\Im\alpha_A{W^{s}_A}+\frac{2k^2\ell}{\kappa^2 \rho}\pqty{\beta_0{W_{3}^{0}}\!\!-\!\!\beta_3{W_{0}^{0}}\!+\!\beta_2{W_{2}^{0}}}\!\!\!\!
    \\
    &-\frac{2k_zk}{\kappa}\frac{Z_\ell^{\prime}}{Z_\ell}\pqty{\beta_0W_{0}^{0}-\beta_2W_{2}^{0}+\beta_3W_{3}^{0}+\beta_1W_{1}^{0}},\!\!\!\!
\\
    \!\!F_{z}=&\sum_{A,s}\pqty{2k_z\Im\alpha_A+{ks}\beta_A}{W^{s}_A}.\!\!
    \end{split}
\end{equation*}%
Note that the chiral force (assuming reciprocal particles $\alpha_2=0$) are only the components which feature $\Re(\alpha_3)$, $\Im(\alpha_3)$, $\beta_0$ and $\beta_1$, i.e.,
\begin{equation*}%
    \begin{split}
    F_{\rho}={}& 2\sum_{s}\kappa\frac{\harmol{\ell-s}'}{\harmol{\ell-s}}\Re\alpha_3{W^{s}_3}
    ,\\
    F_{\varphi}={}& 2\sum_{s}\frac{\ell-s}{\rho}\Im\alpha_3{W^{s}_3}+\frac{2k^2\ell}{\kappa^2 \rho}\pqty{\beta_0{W_{3}^{0}}+\beta_1{W_{2}^{0}}}
    \\
    &-\frac{2k_zk}{\kappa}\frac{Z_\ell^{\prime}}{Z_\ell}\pqty{\beta_0W_{0}^{0}+\beta_1W_{1}^{0}},
\\
    F_{z}={}& 2k_z\Im\alpha_3{W_3}+\sum_{s}k{s}(\beta_0{W^{s}_0}+\beta_1{W^{s}_1}).
    \end{split}
\end{equation*}%

For a visual representation of the different force terms, the total force can still be expressed using the concept of a basis $ \vc{F}=\sum_i \lambda_i \vc{V}\!_i$ as introduced in \cref{eq:basis_concept}. It is then very instructive to plot the twelve different basis vector fields $\vc{V}\!_i$ as given in \cref{tab:basis}. These are the twelve plots shown in the bottom rows of \cref{fig:circular_forces}, and they are entirely determined by the fields of the circularly polarised mode $(\ell=1, n=0)$ while being independent of the particle. To supplement this figure, we also include a Python script \cite{Golat2023} with interactive Jupyter notebooks capable of reproducing these twelve plots for any material parameters and for any mode. 

Examining these basis vector fields $\vc{V}\!_i$, one can deduce that small particles (with negligible recoil, and therefore ignoring the last row) near the fibre will experience a total energy gradient force $\grad W_0$ attracting all particles towards the surface of the fibre, added to a helicity gradient force $\grad W_3$ which attracts one enantiomer towards the fibre, while repelling the other one away from it, exactly as was the case of an evanescent wave described earlier. This is due to the circular polarisation of the mode having a large helicity density in the fibre, which radially decays away, creating a gradient in the helicity. Added to these gradient forces are possible longitudinal ($z$-directed)  pressure forces, as shown in $\vc{p}_0$ and $\vc{p}_3$ in \cref{fig:circular_forces} which are relevant for particles with $\Im(\alpha_\text{e}+\alpha_\text{m})\neq0$ and $\Im\alpha_\text{c}\neq0$, respectively. This can be easily deduced by looking at the coefficients $\lambda_i$ that multiply the relevant basis functions in \cref{tab:basis}. These gradient and pressure forces are completely equivalent to the in-plane separation by the evanescent wave (exactly as in \cref{fig:in-plane} particles b and c), with the only difference being the geometry of the surface to which the particles will stick.  For example, a particle with radius \SI{300}{nm}, relative permittivity $\varepsilon_p=\num{2.5}$, permeability $\mu_p=\num{1}$ and chiral parameter $\abs{\varkappa}=\num{.5}$ (considered in \citet{Li2021}) would feel a chiral force of $\SI{205}{\femto\newton\per\milli\watt}$ near a silicon nitride fibre that guides the fundamental circular mode with wavelength of $\SI{1310}{\nano\meter}$ and radius $\SI{240}{\nano\meter}$. In this case, the chiral forces would be strong enough to separate enantiomers in order of seconds using just $\si{\milli\watt}$ of power (taking the same assumptions as in \citet{Martinez2023}). Therefore, this would be a viable configuration for separating small enantiomers. The provided code \cite{Golat2023} can calculate and plot the force fields for these and any other particle parameters. 
For bigger particles, the recoil forces become stronger. Looking at the last row in \cref{fig:circular_forces}, corresponding to the recoil forces, we can see there is a strong azimuthal component in both the achiral recoil force $k \omega \vc{S}_3 = k \Re{\vc{\varPi}}/c$ and the chiral spin recoil force $k \omega \vc{S}_0 = k \omega (\vc{S}_\text{e}+\vc{S}_\text{m})$ which would suggest a possible enantiomer sorting mechanism in this circular mode. In practice, however, the particle-dependent polarizabilities ($\lambda_i$) multiplying the achiral basis are much stronger than the chiral one, and so the achiral response dominates, making this circular mode not well suited for recoil force sorting. To achieve strong chiral recoil forces that are not overtaken by achiral forces, one must rely on mode combinations to synthesise linearly polarised modes, as shown later. 


\begin{figure*}[h!t]
    \centering
    \includegraphics[width=.85\linewidth]{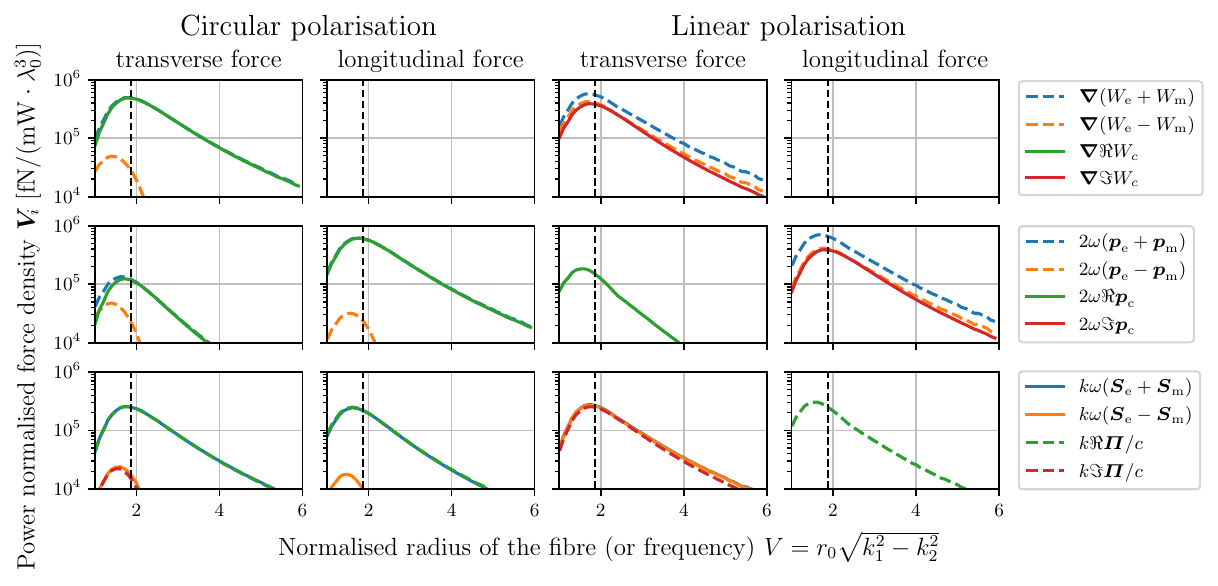}
    \caption{Maximum force densities outside the fibre normalised by the power in units of $\si{\femto\newton\per\milli\watt}\lambda_0^{-3}$ for the fundamental circular polarisation modes with $\ell=1$. Solid lines represent chiral forces, while dashed lines represent achiral forces. Note that these curves are invariant in these units under a change of wavelength or power. The black dashed line represents the optimal radius of the fibre which maximises the chiral gradient force (for CP) and the spin recoil (for LP) (see \cref{eq:opt_rad}). }
    \label{fig:radius}
\end{figure*}%
\begin{figure*}[!p]
    \centering
    \textbf{Fundamental quasi transverse electric (TE) mode of the fibre $\ell=1$, $n=0$}
    \par\medskip
    \includegraphics[width=.40\linewidth]{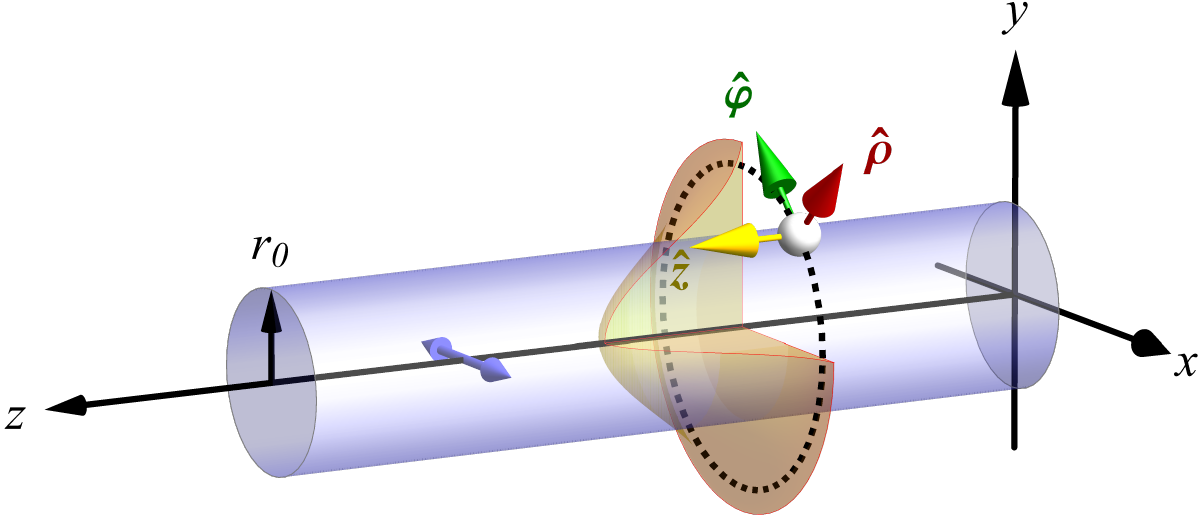}
    \centering
    \includegraphics[width=.96\linewidth]{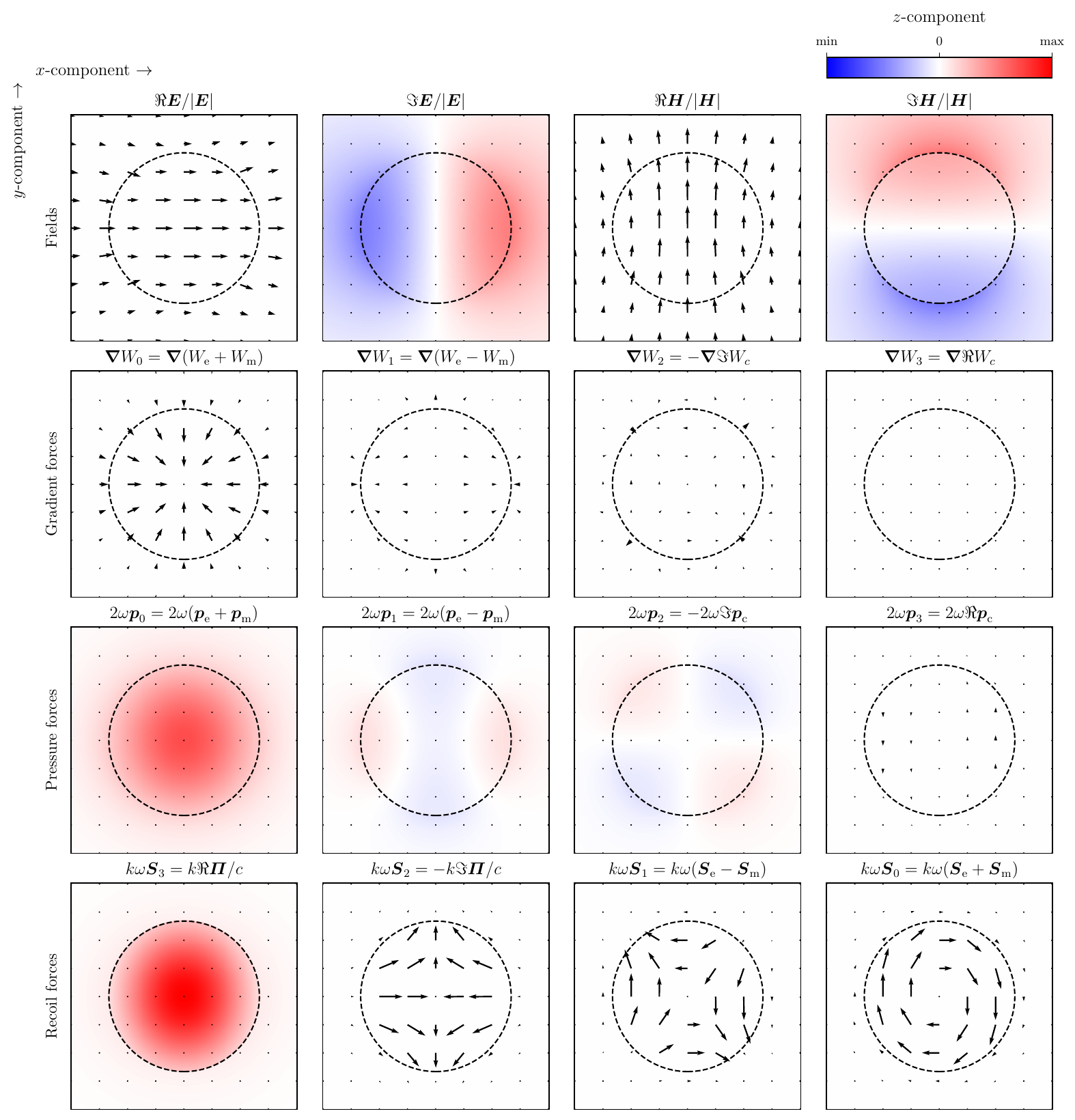}
    \caption{Linear force density basis for optical force around a dielectric fibre. Arrow maps represent transverse components, while colour maps represent longitudinal components of vector quantities plotted in a cross-sectional view of the fibre. 
    This force basis plot is independent of wavelength. It depends on the material of the fibre SiN ($\varepsilon_1=4.3$), the surrounding medium water ($\varepsilon_2=1.7$) and the radius chosen to be $r_0=\num{0.18}\,\lambda_0$.}
    \label{fig:linear_forces}
\end{figure*}%

The cylindrical fibre can support linear combinations of eigenmodes $\vc{E}_{\ell,n}$ \cref{eq:cyl_sol} with complex coefficients $a_{\ell,n}$
\begin{equation}\label{eq:linear_combo}
    \vc{E}=\sum_{\ell}\sum_{n}a_{\ell,n}\vc{E}_{\ell,n}\,,
\end{equation}
and similarly for the magnetic field. Modes with different $\ell$ and $n$, unlike the mode components with different $s$, are not linear in the quadratic observables. Therefore, energy densities will not be simply the sum of individual energy densities. 
\begin{equation*}
    W_A\neq\sum_{\ell}\sum_{n}\sum_{s}W^{\ell,n,s}_A\,.
\end{equation*} Instead, there will also be interference cross terms. For example, take the electric energy density
\begin{equation*}
    W_\text{e}=\frac{\varepsilon}{4}\sum_{\ell,\ell'}\sum_{n,n'}a_{\ell',n'}^*a_{\ell,n}\vc{E}^*_{\ell',n'}\vdot\vc{E}_{\ell,n}\,.
\end{equation*}
where the summand is guaranteed to be real only if $\ell'=\ell$ and $n'=n$, and in that case it would be the electric energy density of that particular mode $W_\text{e}^{\ell,n}$. Other terms will be generally complex, and we only get a real energy density once we perform the sum. Let us, however, label these, in general complex, interference terms $W_A^{\ell,\ell',n,n',s}$, so that, for instance, if $A=\text{e}$ (electric energy) we have
\begin{equation*}
    W^{\ell,\ell',n,n',s}_\text{e}=\frac{\varepsilon}{4}a_{\ell',n'}^*a_{\ell,n}{E}^*_{\ell',n',s}{E}_{\ell,n,s}\,,
\end{equation*}
where $s$ again labels the component in the spin basis. Similarly, we can define them for $A=0,1,2,3$. Canonical momenta and gradients of energy densities will be, in this most general case, as follows:
\begin{equation}
\begin{split}
    \vc{p}_A&=\sum_{\ell,n,s}\Re(n^{\ell,n,s}_A\,\hbar\vc{k}_{\ell,n,s})\,,
\\
    {\grad{W_A}}/{(2\omega)}&=-\sum_{\ell,n,s}\Im(n^{\ell,n,s}_A\,\hbar\vc{k}_{\ell,n,s})\,,  
\end{split}
\end{equation}
where the complex weights in these weighted averages
\begin{equation}
    n^{\ell,n,s}_A=\sum_{\ell',n'}W_A^{\ell,\ell',n,n',s}/(\hbar\omega)
\end{equation}
can be thought of as modified number densities of photons with \emph{complex} momentum $\hbar\vc{k}_{\ell,n,s}$ of the species $A$, but taking into account the interference with all the other eigenmodes. This quantity is, in general, complex and its phase will redistribute (\emph{rotate}) between how much of $\vc{k}_{\ell,n,s}'$ and $\vc{k}_{\ell,n,s}''$ contributes to the gradient and pressure forces. In theory, one can use these interferences to design gradient and pressure forces of arbitrary directions and amplitudes by controlling coefficients $a_{\ell,n}$ in \cref{eq:linear_combo}. However, in practice, we don't have such fine experimental control over these coefficients. 

An example of an experimentally easy-to-obtain mode superposition is that of a linear polarisation. Two orthogonal linear polarisation modes are usually called the quasi-transverse electric (TE) and quasi-transverse magnetic (TM) modes. These modes are obtained in general as the superpositions
\begin{equation}
    \vc{E}_{\ell,n}^\text{TM}=(\vc{E}_{\ell,n}+\vc{E}_{-\ell,n})\,,\quad
    \vc{E}_{\ell,n}^\text{TE}=\ii(\vc{E}_{\ell,n}-\vc{E}_{-\ell,n}).
\end{equation}
The fields of the linearly polarised TE mode with $\ell=1$, $n=0$ can be seen in the top row of \cref{fig:linear_forces}. In this case, the two modes TE and TM are degenerate in the sense that their fields are identical, just rotated by an angle $\pi/2$, and the same is true for all the force densities.

To visualise the possible forces on this TE mode, we once again plot the twelve different basis vector fields $\vc{V}\!_i$ from \cref{tab:basis}, shown in \cref{fig:linear_forces}. In this linear polarisation, the gradient and pressure forces are clearly dominated by the non-chiral total energy gradient $\grad W_0$ and momentum $\vc{p}_0$, hence this mode is not well suited to smaller molecules in which the gradient and pressure forces dominate. Instead, for larger particles on which the recoil forces dominate, this linearly polarised mode shows more promise. Indeed, looking at the recoil terms (last row in \cref{fig:linear_forces}) one can see an achiral Poynting recoil $k \omega \vc{S}_3 = k \Re{\vc{\varPi}}/c$ that is mostly longitudinal ($z$-directed), while there is a strong in-plane azimuthal spin angular momentum recoil force $k \omega \vc{S}_0 = k \omega (\vc{S}_\text{e} + \vc{S}_\text{m})$ pointing around the fibre. This means that opposite enantiomers will be pushed in opposite directions around the fibre, resulting in a potential lateral sorting mechanism. This is caused by the transverse spin, and is exactly equivalent to the lateral sorting in linearly polarised evanescent waves from the previous section. Interestingly, this force competes with an achiral force proportional to the flux of reactive power $-k \omega \vc{S}_2 = k \Im\vc\varPi/c$ which depends on the coefficient $\beta_2\sim\Im(\alpha_\text{e}^*\alpha_\text{m})$ and which is present for any particle that is breaking dual symmetry (i.e. a dominant electric or magnetic response). For the same particle as before ($a=\SI{300}{nm}$, $\varepsilon_p=\num{2.5}$, $\mu_p=\num{1}$, and $\abs{\varkappa}=\num{.5}$) 
and linear polarisation with $\SI{1310}{\nano\meter}$ in a silicon nitride fibre with $r_0=\SI{240}{\nano\meter}$ the maximal chiral force on the particle would be $\SI{218}{\femto\newton\per\milli\watt}$, which would again make it possible to separate these particles in order of seconds. 

Finally, it is interesting to study the effect that the fibre radius has on the different force terms. To this end, \cref{fig:radius} plots the maximum magnitude of each of the twelve force term basis $\vc{V}\!_i$ in both the circularly polarised and linearly polarised fibre modes as a function of normalised radius. One can see that there is clearly an optimal fibre radius-to-wavelength ratio. This optimum is the sweet spot between the radius being too large (hence mode not interacting with the particles outside) and too small (hence mode fields very spread outside, with low gradients). In the limit where the fibre is in a medium for which $n_2=0$ this radius coincides with the point at which the mode with $\ell=0$ starts. The condition for a single-mode fibre operation is $k_0r_0\sqrt{n_1^2-n_2^2}<\mathrm{j}_{0,1}$\cite{Jurgensen1975}, where $\mathrm{j}_{0,1}$ is the first root of a Bessel function $\bessel{0}$ with the approximate value of $2.40483$. While the radius at which the fibre can support additional modes depends on $n_2$, the optimal radius does not, leading to a simple expression for the optimal radius to wavelength ratio for any value of $n_2$:
\begin{equation}\label{eq:opt_rad}
    \frac{r_0}{\lambda_0}=\frac{\mathrm{j}_{0,1}}{2\pi n_1},
\end{equation}
which only depends on the wavelength and the refractive index of the material of the fibre.

\section{Conclusions}
The total optical force acting on a general dipolar particle is a relatively large analytical expression that can be written in many alternative but equivalent ways. We provided a clear view of the different alternatives used in the literature and introduced our own alternative that classifies forces based on the symmetries broken by the particle and the fields. The chiral forces that separate enantiomers rely on particles and fields that break parity---but other forces can separate, for instance, electric from magnetic particles, which break duality symmetry. The concept of a force basis, relying on twelve force fields, each with its own particle-dependent coefficient, was introduced---and their manifestation in evanescent waves and nanofibre modes was exemplified. We also developed very concise analytical expressions for the optical forces in such modes---including the case of arbitrary nanofibre modes and their combination. These analytical expressions suggest that the separation of bigger enantiomers, where the recoil force dominates, should rely on transverse-spin-based lateral forces requiring linearly polarised modes, while separation of smaller (molecule-sized) enantiomers should rely on gradient and pressure forces---so the use of circularly polarised modes provides a helicity gradient to attract or repel opposite enantiomers towards or away from the nanofibre. An insight that seems to be also true in the case of a rectangular waveguide \cite{Martinez2023}. An optimal radius was found for maximising forces in nanofibres, which depends only on their material and wavelength. Finally, we provide an interactive Python script \cite{Golat2023} that can reproduce all \cref{fig:circular_forces,fig:linear_forces,fig:radius} for an arbitrary choice of parameters and modes.
We hope that our theoretical work provides guidance and clarity for the design of future experimental attempts at optical separation of chiral enantiomers near waveguides, which is of enormous practical importance in the pharmaceutical domain.

\begin{acknowledgments}
We would like to acknowledge the financial support from the European Innovation Council (HORIZON-EIC) Pathfinder CHIRALFORCE 101046961.
\end{acknowledgments}

\appendix
\section{Dynamic polarisabilities of bi-isotropic dipolar particle}\label{app:polar}
For spherical particles of radius $a$, and relative material properties ($\varepsilon_p$, $\mu_p$, $\varkappa$), where $\varkappa$ is the material chirality, in a non-chiral background medium with relative permittivity and permeability ($\varepsilon_m$, $\mu_m$), one can use the Clausius-Mossotti expressions for static polarisabilities 
\begin{equation}%
\begin{split}%
\alpha_\text{0e} & =4 \pi a^3 \frac{\left(\varepsilon_p-\varepsilon_m\right)\left(\mu_p+2 \mu_m\right)-\varkappa^2}{\left(\varepsilon_p+2 \varepsilon_m\right)\left(\mu_p+2 \mu_m\right)-\varkappa^2} \\
\alpha_\text{0m}& =4 \pi a^3 \frac{\left(\varepsilon_p+2 \varepsilon_m\right)\left(\mu_p-\mu_m\right)-\varkappa^2}{\left(\varepsilon_p+2 \varepsilon_m\right)\left(\mu_p+2 \mu_m\right)-\varkappa^2} \\
\alpha_\text{0c} & =12 \pi a^3 \frac{\varkappa}{\left(\varepsilon_p+2 \varepsilon_m\right)\left(\mu_p+2 \mu_m\right)-\varkappa^2}\label{eq:chiralCMpol}
\end{split}%
\end{equation}%
As shown by \citet{Sersic2011}, in order for the polarisabilities to satisfy energy conservation, one has to apply the tensor radiative correction by adding radiation damping 
\begin{equation}\label{eq:radiative_correction}
    \tens{\alpha}^{-1}=\tens{\alpha}^{-1}_0-\ii\frac{k^3}{6\pi}\tens{I}\,,
\end{equation}
where $\tens{\alpha}$ is the full $6\times6$ square matrices appearing in \cref{eq:generaltensorpolarisability} and representing all the polarisabilities.
While the importance of radiative corrections is not new, to our knowledge, it has not been always applied correctly in the chiral case. Sometimes the correction \cref{eq:corrections} is applied only to the electric and magnetic polarisabilities, assuming the chiral polarisability to be zero, while the static chiral polarisability from \cref{eq:chiralCMpol} is uncorrected. For a bi-isotropic dipolar particle, the radiative corrections coming from \cref{eq:radiative_correction} are as follows:
\begin{equation}\label{eq:corrections}
\begin{split}
    \!\!\!\alpha_\text{e}&=\frac{\alpha_\text{0e}
    -\ii\frac{ k^3}{6 \pi}(\alpha_\text{0c}^2-\alpha_\text{0e}\alpha_\text{0m})}{1+\qty(\frac{k^3}{6 \pi})^2\qty(\alpha_\text{0c}^2-\alpha_\text{0e} \alpha_\text{0m})-\ii \frac{k^3}{6 \pi}\qty(\alpha_\text{0e}+\alpha_\text{0m})},\!\!\!\\
    \!\!\!\alpha_\text{m}&=\frac{\alpha_\text{0m}
    -\ii\frac{ k^3}{6 \pi}(\alpha_\text{0c}^2-\alpha_\text{0e}\alpha_\text{0m})}{1+\qty(\frac{k^3}{6 \pi})^2\qty(\alpha_\text{0c}^2-\alpha_\text{0e} \alpha_\text{0m})-\ii \frac{k^3}{6 \pi}\qty(\alpha_\text{0e}+\alpha_\text{0m})},\!\!\!\\
    \!\!\!\alpha_\text{c}&=\frac{\alpha_\text{0c}
    }{1+\qty(\frac{k^3}{6 \pi})^2\qty(\alpha_\text{0c}^2-\alpha_\text{0e} \alpha_\text{0m})-\ii \frac{k^3}{6 \pi}\qty(\alpha_\text{0e}+\alpha_\text{0m})}.\!\!\!
\end{split}
\end{equation}

\section{Fluxes of energy and helicity densities}\label{app:fluxes}
Interestingly, the four \emph{spin-like} quantities, $\vc{S}\!_A$, in \cref{eq:spins} can be seen as the fluxes of the energy densities $W\!\!_A$ in \cref{eq:energies}. Starting with the real Poynting vector $\Re{\vc\varPi}=\omega c\vc{S}_3$, the Poynting theorem (continuity equation for active power) relates the time derivative of the total active energy density $W_0$ to $\Re{\vc\varPi}$ which can be interpreted as the flow of active power \cite{Jackson1998}. Somewhat less known is the recently proposed complex Poynting theorem \cite{Jackson1998,Kaiser2016} whose real part gives the usual time-averaged Poynting theorem for a time-harmonic field,
\begin{equation}
    -\omega c \div\vc{S}_3=\frac{1}{2}\Re(\vc{E}^\ast\vdot\vc{J}),
\end{equation}
where the term with $W_0$ vanishes since the average energy density does not change with time. The imaginary part relates the reactive or stored energy density $W_1$ and an alternating 
flow of reactive power $\Im{\vc\varPi}=-\omega c\vc{S}_2$, which in our notation reads:
\begin{equation}\label{eq:imaginary_poynting}
    2\omega W_1-\omega c \div\vc{S}_2=\frac{1}{2}\Im(\vc{E}^\ast\vdot\vc{J}).
\end{equation}
where the term on the right-hand side is the reactive power. Similarly, it is a well-known fact that the conservation and flow of integrated optical helicity (the continuity equation) relates
helicity density $\mathfrak{S}=W_3/\omega$ to the spin (helicity flux) density $\vc{S}_0$ \cite{Cameron2012a,Bliokh2013}. For time-averaged quantities, the $W_3$ will be again missing like in the case of the real Poynting theorem because the time-averaged helicity will be conserved and we will get
\begin{equation}
    -\omega c \div\vc{S}_0=\frac{1}{4}\Im(\eta\vc{H}^\ast\vdot\vc{J})
\end{equation}
notice the appearance of pseudoscalar quantity in units of power on the right-hand side, which seems to be a chiral equivalent of active power.
A similar relationship to \cref{eq:imaginary_poynting} then exists between the remaining two quantities, suggesting that $\vc{S}_1$ is the flow of $W_2$
\begin{equation}
    2\omega W_2-\omega c \div\vc{S}_1=\frac{1}{4}\Im(\eta\vc{H}^\ast\vdot\vc{J}).
\end{equation}
\section{Chiral energy, canonical momentum and spin angular momentum densities}\label{app:chiraldensities}
One can show that chiral energy density 
$W_3$, chiral momentum $\vc{p}_3$ and chiral spin angular momentum density $\vc{S}_3$ are just differences between energies, momenta and spins carried by the right- and left-handed fields. In order to do that one can write the angular spectrum decomposition of the electric field to separate it into positive and negative helicity components
\begin{equation}
\begin{split}
    \vc{E}(\vc{r})&=\!\!\iiint\tilde{\vc{E}}(\vc{k})\ee^{\ii\vc{k}\vdot\vc{r}}\dd[3]k\\
    &=\!\!\iiint[\tilde{\vc{E}}_+(\vc{k})\uv{e}_+(\vc{k})+\tilde{\vc{E}}_-(\vc{k})\uv{e}_-(\vc{k})]\ee^{\ii\vc{k}\vdot\vc{r}}\dd[3]k\\
    &=\vc{E}_+(\vc{r})+\vc{E}_-(\vc{r})\,,
\end{split}
\end{equation}
where $\uv{e}_\pm(\vc{k})$ are the circularly polarised basis vectors for each plane wave with wave-vector $\vc{k}$. We can find the magnetic fields using the Maxwell-Faraday equation $\curl{\vc{E}}=\ii\omega\mu\vc{H}$ leading to 
\begin{equation}
\begin{split}
    \!\!\vc{H}(\vc{r})&=\frac{1}{\eta}\!\iiint\uv{k}\cp\tilde{\vc{E}}(\vc{k})\ee^{\ii\vc{k}\vdot\vc{r}}\dd[3]k\\
    &=\frac{1}{\eta}\!\iiint[-\ii\tilde{\vc{E}}_+\uv{e}_+(\vc{k})+\ii\tilde{\vc{E}}_-\uv{e}_-(\vc{k})]\ee^{\ii\vc{k}\vdot\vc{r}}\dd[3]k\\
    &=\vc{H}_+(\vc{r})+\vc{H}_-(\vc{r})=-\frac{\ii}{\eta}[\vc{E}_+(\vc{r})-\vc{E}_-(\vc{r})]\,.\!\!\!\!
    \end{split}
\end{equation}
Using the definition of $W_3$ one can show that
\begin{align*}
    W_3={}&\Re W_\text{c}=\frac{1}{4 c}\Im\bqty{\vc{H}^\ast\vdot\vc{E}-\vc{E}^\ast\vdot\vc{H}}\\
    ={}&\frac{1}{4\omega }\Re\{\varepsilon[\vc{E}_++\vc{E}_-]^\ast\vdot[\vc{E}_+-\vc{E}_-]\\
    &+\varepsilon[\vc{E}_+-\vc{E}_-]^\ast\vdot[\vc{E}_++\vc{E}_-]\}\\
    ={}&\frac{1}{4\omega}\pqty{\varepsilon\abs{\vc{E}_+}^2-\varepsilon\abs{\vc{E}_-}^2}=W_+-W_-\,,
\end{align*}
where $W_\pm$ is the energy density carried by the right- or left-handed component of the field. The same can be done for the canonical momentum density
\begin{align*}
    \vc{p}_3={}&\Re\vc{p}_\text{c}=\frac{1}{4\omega c}\Re\bqty{\vc{E}^\ast\vdot(\grad)\vc{H}-\vc{H}^\ast\vdot(\grad)\vc{E}}\\
    ={}&\frac{1}{4\omega }\Im\{\varepsilon[\vc{E}_++\vc{E}_-]^\ast\vdot(\grad)[\vc{E}_+-\vc{E}_-]\\
    &+\varepsilon[\vc{E}_+-\vc{E}_-]^\ast\vdot(\grad)[\vc{E}_++\vc{E}_-]\}\\
    ={}&\frac{1}{4\omega}\Im\bqty{\varepsilon\vc{E}_+^*\vdot(\grad)\vc{E}_+-\varepsilon\vc{E}_-^\ast\vdot(\grad)\vc{E}_-}=\vc{p}_+-\vc{p}_-\,,
\end{align*}
and for the spin angular momentum density that is related to the Poynting vector
\begin{align*}
    \vc{S}_3={}&\frac{1}{\omega c}\Re\vc{\varPi}=\frac{1}{4\omega c}\Re\bqty{\vc{E}^\ast\cp\vc{H}-\vc{H}^\ast\cp\vc{E}}\\
    ={}&\frac{1}{4\omega }\Im\{\varepsilon[\vc{E}_++\vc{E}_-]^\ast\cp[\vc{E}_+-\vc{E}_-]\\
    &+\varepsilon[\vc{E}_+-\vc{E}_-]^\ast\cp[\vc{E}_++\vc{E}_-]\}\\
    ={}&\frac{1}{4\omega}\Im\bqty{\varepsilon\vc{E}_+^*\cp\vc{E}_+-\varepsilon\vc{E}_-^\ast\cp\vc{E}_-}=\vc{S}_+-\vc{S}_-\,.
\end{align*}

\section{Boundary conditions for a cylindrical dielectric fibre}\label{app:bondary}
At the interface between the dielectric fibre and the surrounding material, the tangent components of the electric field $E_z$, $E_\varphi$ and magnetic field $H_z$, $H_\varphi$ have to be continuous.
These boundary conditions can be written as an eigenvalue problem 
\begin{equation}
    \vc{A}_{\ell,n}\vc{v}_{\ell,n}=0,
\end{equation}
where $\vc{v}_{\ell,n}=(A_1,B_1,A_2,B_2)^\intercal$ represents the amplitudes both outside and inside the fibre and the boundary condition matrix as a function of 
$(k_z,\omega, r_0, \varepsilon_1,\varepsilon_2)$ is as follows:
\begin{equation*}
    \vc{A}_{\ell,n}=\!\!\pmqty{
n_2 \bessel{\ell} & 0 & - n_1\hankel{\ell} & 0 \\
\!\! \frac{\ell n_2k_z}{\kappa_1^2 r_0} \bessel{\ell}\!\! &  \frac{\ii n_2k_1}{\kappa_1} \bessel{\ell}^{\prime} & - \frac{\ell n_1k_z}{\kappa_2^2 r_0} \hankel{\ell}\!\! & -  \frac{\ii n_1k_2}{\kappa_2} \hankel[(1)\prime]{\ell}\!\!\!\! \\
0 & \bessel{\ell} & 0 & -\hankel{\ell} \\
-\text{i} \frac{k_1}{\kappa_1} \bessel{\ell}^{\prime} & \frac{\ell k_z}{\kappa_1^2 r_0} \bessel{\ell} & \text{i} \frac{k_2}{\kappa_2} \hankel[(1)\prime]{\ell}\!\! & -\frac{\ell k_z}{\kappa_2^2 r_0} \hankel{\ell}
},
\end{equation*}%
where $n_i=\sqrt{\varepsilon_i}$ is the refractive index of medium $i$, $\bessel{\ell}=\bessel{\ell}(\kappa_1\rho)$, $\hankel{\ell}=\hankel{\ell}(\kappa_2\rho)$ are Bessel and Hankel functions of the first kind respectively and the prime represents derivative of the whole argument of a function.
Given the characteristic parameters $(\omega, r_0, \varepsilon_1,\varepsilon_2)$, one can numerically calculate the longitudinal dispersion relation $k_z/k_0$ as a function of $k_0r_0$ for any mode from the transcendental equation $\det\boldsymbol{A}_{\ell,n}(k_z,\omega, r_0, \varepsilon_1,\varepsilon_2)=0$. The solutions of this equation for $\ell\in\{0,\pm1,\pm2\}$ and $n\in\{0,1,2\}$ can be seen in \cref{fig:dispersion} using standard normalised units popular in the literature.

%

\end{document}